\documentclass{aastex631}

\accepted{13th February 2023}
\submitjournal{ApJ}

\usepackage[british]{babel}
\usepackage{amsmath}
\usepackage{graphicx}

\begin{document}

\title{Formulating the r-mode problem for slowly rotating neutron stars}

\author[0000-0001-8550-3843]{Nils Andersson}
\affiliation{Mathematical Sciences and STAG Research Centre, 
             University of Southampton, 
             Southampton SO17 1BJ, 
             United Kingdom}
\email{n.a.andersson@soton.ac.uk}

\author[0000-0002-9439-7701]{Fabian Gittins}
\affiliation{Mathematical Sciences and STAG Research Centre, 
             University of Southampton, 
             Southampton SO17 1BJ, 
             United Kingdom}
\email{f.w.r.gittins@soton.ac.uk}

\begin{abstract}
We  revisit the problem of inertial r-modes in stratified stars, drawing on a more precise description of the composition stratification in a mature neutron star. The results highlight issues with the traditional approach to the problem, leading us to rethink the computational strategy for r-modes of non-barotropic neutron stars. We outline two strategies for dealing with the problem. For moderate to slowly rotating neutron stars the only viable alternative may be to approach the problem numerically from the outset, while a meaningful slow-rotation calculation can be carried out for the fastest known spinning stars (which may be close to being driven unstable by the  emission of gravitational waves).  We demonstrate that the latter approach leads to a problem close, but not identical, to that for barotropic inertial modes. We also suggest that these reformulations of the problem likely resolve the long-standing problem of singular behaviour associated with a co-rotation point in rotating relativistic neutron stars. This issue needs to be resolved in order to guide future  gravitational-wave searches.
\end{abstract}


\section{Motivation}

The inertial r-modes of a spinning neutron star have attracted a fair amount of attention. Theorists have explored the precise nature of the r-modes and how they depend on the complex physics of a neutron star interior, while observers have tried to establish the presence of the predicted r-mode signature in observational data. Much of this interest stems from the discovery (now a quarter of a century ago!) that the r-modes may be driven unstable by the emission of gravitational waves \citep{1998ApJ...502..708A, 1998ApJ...502..714F,1998PhRvL..80.4843L,1999ApJ...510..846A}. The r-mode instability may limit the spin-up of accreting neutron stars in low-mass x-ray binaries \citep{1999ApJ...516..307A}, providing a natural explanation for the apparent absence of sub-millisecond radio pulsars.
The mechanism may also lead to the emission of detectable gravitational waves from newly born neutron stars \citep{1998PhRvD..58h4020O}, mainly through the current multipoles associated with the induced fluid motion. This has motivated a sequence of observational gravitational-wave papers \citep{2010ApJ...722.1504A,2015ApJ...813...39A, 2020ApJ...895...11F,2021ApJ...921...80A, 2021ApJ...922...71A,2022PhRvD.105b2002A, 2022PhRvD.106d2003A,2022ApJ...929L..19C}, so far mainly setting upper limits on the attainable r-mode amplitude. There have also been tantalizing hints of r-mode oscillations in the the x-ray emission from two fast spinning, accreting neutron stars 
\citep{2014ApJ...784...72S,2014ApJ...793L..38S}, see also \citep{2014MNRAS.442.3037L,2014MNRAS.442.1786A}, but these results are far from conclusive.

The need to understand the impact of different aspects of neutron star physics---ranging from the main dissipation channels (shear and bulk viscosity), the state of matter (superfluid mutual friction and the interfaces with the elastic crust) and the role of the star's magnetic field---led to a flurry of activity following the original instability discovery. This work is summarised in early review articles, see \citet{2001IJMPD..10..381A} and \citet{2003CQGra..20R.105A}, with more recent additions like \citet{2015PASA...32...34L} and \citet{2021arXiv210403137H} providing a  mature perspective. While many aspects of the problem are fairly well understood some vexing issues remain. Arguably, the most important of these issues relates to the r-modes in relativity. 

In order to make robust predictions for (say) the r-mode frequency in a neutron star we need to involve a realistic matter equation of state. This, in turn, requires a general relativistic mode calculation. This problem has not---in our view---yet been solved in a satisfactory fashion. Let us explain. The most important contributions to the discussion, from the initial relativistic inertial-mode calculations by \citet{2000PhRvD..63b4019L,2003PhRvD..68l4010L} and \citet{2003MNRAS.339.1170R} through to the more recent work for real equations of state by \citet{2015PhRvD..91b4001I},  assumes that the matter is barotropic. However, this is not expected to be  a realistic assumption. Instead, as established by \citet{1992ApJ...395..240R}, the stratification associated with internal composition gradients makes the problem non-barotropic. This seems to complicate the r-mode calculation. In fact, the problem appears to become singular leading to a continuous spectrum   \citep{1998MNRAS.293...49K,1999ApJ...520..788K,1999MNRAS.308..745B}. The implications of this are not well understood, but it seems reasonable to argue that the continuous spectrum arises because of simplifying assumptions introduced in the analysis. Adopting this view, the question becomes how we can regularise the problem. While different strategies have been proposed, see  
\cite{2004CQGra..21.4661L,2002ApJ...567.1112Y,2005MNRAS.363..121P} and the recent effort from \cite{2021arXiv211201171K, 2022Univ....8..542K}, it is probably fair to suggest that the issue has not yet been resolved---at least not completely. This motivates us to return to the problem.

Our aim is to formulate the r-mode problem for neutron stars stratified by composition gradients from first principles. This forces us to consider how the non-barotropic aspects arise and how this affects the fluid perturbation equations. The key point is that the matter composition may be considered frozen provided the relevant nuclear reactions are slow compared to the dynamics. The argument was already outlined by \cite{2019MNRAS.489.4043A}, although in that case the main focus was on the composition g-modes. Here we take one step further by framing the discussion in the context of a realistic matter model. This leads to an important---in hindsight probably obvious---insight. While realistic neutron star models are likely to have non-barotropic high-density cores they will always have barotropic---basically because the matter is composed of single nuclei at lower densities---outer layers. This has two immediate repercussions. First,  the standard assumption of a constant adiabatic index ($\Gamma_1$) for the perturbations is never appropriate. Second, none of the existing r-mode calculations actually solve the problem we should be considering. We have have to rethink how we approach the problem. 

Ultimately, the implications of our new perspective on the problem will depend on the extent to which numerical mode results differ from existing ones. This problem will not be solved here. In this first paper we are mainly interested in the qualitative aspects and so we introduce a number of approximations that sacrifice accuracy for clarity. We want to make the key points as transparent as possible in order to motivate renewed effort in several directions. First and foremost, we need relativistic mode calculations for true equations of state in order to inform future gravitational-wave searches. It may well be that the current models are ``good enough'' but---as we are arguing for changes to the formulation of the problem---this is by no means guaranteed. Second, we need to revisit the (technically challenging) problem of nonlinear mode-coupling and saturation of the r-mode instability \citep{2001PhRvD..65b4001S,2003ApJ...591.1129A,2004PhRvD..70l4017B,2004PhRvD..70l1501B,2005PhRvD..71f4029B,2007PhRvD..76f4019B,2013ApJ...778....9B,2009PhRvD..79j4003B}. Available mode-coupling results relate to barotropic models and hence may change when we add realism to the discussion. Third, the precise nature of the 
r-modes impact on the  gravito-magnetic interaction and the dynamical tides in spinning neutron star binaries \citep{2007PhRvD..75d4001F}. Again, this is a problem that has so far only been explored for barotropic stellar models \citep{2017PhRvD..96h3005X, 2020PhRvD.101j4028P, 2020PhRvD.102j4005P, 2021PhRvD.103f3020M}. 

It appears that there is quite a lot of work to get on with, so let us get started.

\section{Low-frequency oscillations of a rotating star}
\label{lowfreq}

 In general, the oscillation modes of a rotating star belongs to one of two  categories: modes that exist (have a finite eigenfrequency) already in a non-rotating star and modes that owe their existence to the rotation (the Coriolis force). Our main interest here is in the latter class, collectively referred to as inertial modes, but the story will not be complete unless we also touch upon the former. The argument at the centre of our discussion also impacts on the gravity g-modes, which are present if the star is stably stratified, either in terms of an entropy gradient or a varying composition. This is a non-trivial issue. Work on main-sequence stars demonstrates that the impact of stratification may vary, with global  oscillation modes having different character in different parts of a star \citep{1987MNRAS.224..513L}. An illustrative example concerns slowly pulsating B stars \citep{2006MNRAS.365..677L}.

Work on the oscillations of rotating stars is complicated by the fact that the Coriolis force breaks the spherical symmetry of the non-rotating case, leading to a coupling of the angular harmonics traditionally used to represent the modes. This coupling is particularly significant for modes with frequency comparable to or smaller than the star's rotation frequency. Higher frequency modes may be approached ``perturbatively'', adding rotational corrections to the modes of a non-rotating star, but the low-frequency problem is intricate. This issue is not at all new. It was recognised already in the seminal work on rotating ellipoids by  \cite{1889RSPTA.180..187B} (for a  modern version of the calculation, see \cite{1999PhRvD..59d4009L}). It is also well known from work on waves in shallow ocean/atmospheres, mainly focused on weather and climate studies. 
The r-modes---the main focus of our attention---are in fact analogous to the Rossby waves from the shallow water problem (see, for example, \citet{1968RSPTA.262..511L,2021SSRv..217...15Z}).

For rotating stars, the r-modes were first introduced by 
\cite{1978MNRAS.182..423P}. Their work was followed by a number of detailed studies in the early 1980s  \citep{1981A&A....94..126P, 1982ApJ...256..717S, 1981ApJ...244..299S, 1983A&A...125..193S}. The nature of the problem was laid out in detail, relaxing the Cowling approximation (i.e. allowing for perturbations of the gravitational potential), by \cite{1981Ap&SS..78..483S}. This body of work establishes the main features of the r-modes. They are represented by  retrograde waves in the co-rotating frame of
the star, but appear prograde in the inertial frame. It is precisely this character that makes them unstable to the emission of gravitational waves \citep{1998ApJ...502..708A, 1998ApJ...502..714F}.

The problem is complicated by the fact that the modes are degenerate in barotropic stars, belonging to the broader class of inertial modes (with frequency proportional to the star's angular frequency) \citep{1999ApJ...521..764L}. This changes when the stellar fluid is stably stratified. 
Stratification couples the radial layers in the star, breaking the degeneracy of the modes and allowing for a richer spectrum of r-modes (including radial overtones) \citep{1981A&A....94..126P,1982ApJ...256..717S}. In the context of the r-mode instability, the work by  \cite{2000ApJS..129..353Y} is particularly notable
although they consider either models with  stable  stratification  throughout the
interior or models that are fully convective. However---and this is important---the arguments they put forward for the stratification relate to entropy gradients, which are only expected to be relevant for newly born neutron stars. 

We will set up the problem in such a way that our discussion connects with the inertial-mode analysis of \cite{1999ApJ...521..764L}. This makes sense because one of the questions we want to address involves how the non-barotropic r-modes morph into barotropic inertial modes as the stratification weakens. 

Starting from the velocity perturbations $(\delta v^i)$, we consider the perturbed Euler equation in the rotating frame of the star. We then have (in a coordinate basis, making due distinction between co- and contra-variant components, and using $\delta$ to indicate Eulerian variations) 
\begin{equation}
\partial_t \delta v_i + 2 \epsilon_{ijk}\Omega^j
\delta v^{k}  + { 1 \over \rho} \nabla_i\delta p - { 1 \over \rho^2} \delta \rho \nabla_i p +  \nabla_i \delta \Phi
 = 0 \ ,
\end{equation}
where $p$ is the pressure, $\rho$ is the mass density and $\Phi$ the gravitaitonal potential---along with  the continuity equation
\begin{equation}
\partial_t \delta \rho + \nabla_i (\rho \delta v^i) =  0  
\label{cont}
\end{equation}
and the Poisson equation for the perturbed gravitational potential
\begin{equation}
    \nabla^2 \delta \Phi = 4\pi G \delta \rho \ .
    \label{potens}
\end{equation}

In the static limit (where time variations vanish and the rotation frequency $\Omega^i\to0$) the equations decouple into two sets. First we have
\begin{equation}
  \nabla_i \delta \Phi + { 1 \over \rho} \nabla_i \delta p - { 1 \over \rho^2} \delta \rho \nabla_i p
 = 0 \ ,
 \label{neighbour}
\end{equation}
along with \eqref{potens} and, secondly,
\begin{equation}
\nabla_i (\rho \delta v^i) =  0  \ .
\label{div}
\end{equation}

At this point, \cite{1999ApJ...521..764L} note that the first two equations represent perturbations that take us to a neighbouring equilibrium star. The argument for this is straightforward---a static perturbation of the equation for hydrostatic equilibrium takes us to a new  configuration with pressure $\bar p = p + \delta p$. 
In a slowly rotating star, this solution would pick up rotational corrections at order $\Omega$, which suggests  a solution  such that
\begin{equation}
    [\delta \rho, \delta p, \delta \Phi] = \mathcal O(1) \quad \mathrm{and} \quad \delta v^i = \mathcal O(\Omega) \ .
\end{equation}
This problem is  equivalent to considering the dynamics of the original configuration (albeit for a slight different central density). The dynamical aspects of the problem are contained in the second set of perturbations, represented by equation \eqref{cont} for which we would have
\begin{equation}
    \delta v^i = \mathcal O(1) \quad \mbox{and} \quad [\delta \rho, \delta p, \delta \Phi] = \mathcal O(\Omega) \ .
\end{equation}
This is the  assumption that leads to the inertial modes. Alternatively, as we can always multiply the linearised equations by a constant, we may normalise the Lagrangian displacement associated with the perturbation, in the rotation frame simply given by
\begin{equation}
    \delta v^i = \partial_t \xi^i\ ,
\end{equation}
such that
\begin{equation}
   \xi^i = \mathcal O(1) \Longrightarrow \delta v^i = \mathcal O(\Omega) \quad \mbox{and} \quad   [\delta \rho, \delta p, \delta \Phi] = \mathcal O(\Omega^2) \ .
   \end{equation}
This is the convention we  assume in the following. It is important to appreciate that, regardless of the chosen normalisation, we cannot (completely) determine the density perturbations etc without accounting for the change in shape due to the centrifugal force (which also enters at order $\Omega^2$). This obviously complicates the analysis and it would be natural to turn to numerics. However, in order to understand the nature of the problem we wish to proceed analytically (as far as we can). We  consider the numerical problem in a companion paper \citep{FandA}. In order to make progress analytically, we will study the perturbations to second order in $\Omega/\Omega_0$, where
\begin{equation}
    \Omega_0^2 = {GM\over R^3}\ , 
\end{equation}
However, in the interests of clarity, we will keep the background spherical \citep{1987MNRAS.224..513L}. This is a useful simplification as it removes some of the rotational multipole couplings from the problem. It also  makes ``sense'' since the centrifugal force does not introduce additional oscillation modes. In parts of the discussion we will also, again for clarity, make use of the Cowling approximation (neglect the perturbed gravitational potential, $\delta \Phi$). Both assumptions are relaxed in the companion numerical work.

\subsection{The frozen composition argument}

It is easy to see how the assumption of non-barotropic perturbations upsets the inertial-mode logic. The usual argument introduces the adiabatic index $\Gamma_1$ in such a way that (with $\Delta$ representing Lagrangian variations)
\begin{equation}
    \Delta p = {p\Gamma_1 \over \rho} \Delta \rho \Longrightarrow \delta p = {p\Gamma_1 \over \rho} \delta \rho  + {p\Gamma_1 \over \rho} \xi^i \nabla_i \rho - \xi^i \nabla_i p \ .
    \label{Gamma1}
\end{equation} 
This immediately leads to a conflict with the assumed ordering for the rotating-star perturbations \citep{1999ApJ...521..764L}. 
Since the background is spherical  we must either have 
\begin{equation}
  {p\Gamma_1 \over \rho} {d \rho \over dr} = {dp\over dr} \Longrightarrow \Gamma_1 = {\rho \over p} {dp\over d\rho} = {\rho c_s^2 \over p} \equiv \Gamma \ ,
  \label{Gamma}
\end{equation}
(introducing both the adiabatic index $\Gamma$ and the speed of sound $c_s^2$ for the background configuration)
which would represent a barotropic model, or the ordering of the solution must change in such a way that  
$\xi^r = \mathcal O(\Omega^2)$ to balance the density and pressure perturbations. This then leads to the different starting assumption for non-barotropic models suggested by   \cite{1999ApJ...521..764L} (and eventually brings us to the vexing issue for relativistic r-modes, see \citep{2004CQGra..21.4661L}). However, for realistic neutron star physics, the argument turns out to be a bit more subtle. 

In order to explain the issue we need to explore the physics that give rise to the stratification in a mature neutron star (internal composition gradients) in the first place, leading to $\Gamma_1\neq \Gamma$. The argument draws heavily on the  discussion of g-modes and reactions from \cite{2019MNRAS.489.4043A}. Essentially, once we account for out-of equilibrium nuclear reactions\footnote{Our discussion assumes a star dominated by npe-matter, for which the key reactions are due to the Urca processes. This is the simplest relevant case. The problem will change if (say) hyperons or deconfined quarks are present at high densities. The reaction rates will then be different as may be the outcome for the stratification. Similarly, the state of matter is important. For example, it is known that the composition gradient in a superfluid neutron star core arises not due to an imbalance between neutron and protons, but as a result of the presence of muons \citep{2013PhRvD..88j1302G,2016MNRAS.455.1489P}.}, the perturbed proton fraction $x_\mathrm{p}$ evolves according to (for small deviations from equilibrium---i.e. in the so-called sub-thermal limit  \citep{2018PhRvC..98f5806A})
\begin{equation}
 (\partial_t + v^j \nabla_j ) \Delta x_\mathrm{p} = {\gamma\over n} \Delta \beta \ .
 \label{eqtwo}
\end{equation}
Here $\Delta \beta$ represents the deviation from beta-equilibrium and
  $\gamma$ encodes the (dominant) reaction rate.
Considering $\beta$ a function of $\rho$ and the proton fraction $x_\mathrm{p}$, and assuming that the star is non-rotating (so that $v^i=0$, which is also true in the rotating frame), we  have 
\begin{equation}
\partial_t \Delta \beta = \left( {\partial \beta \over \partial \rho}\right)_{x_\mathrm{p}}  \partial_t \Delta \rho +  \left( {\partial \beta \over \partial x_\mathrm{p}}\right)_{\rho} \partial_t  \Delta x_\mathrm{p} \ ,
\end{equation}
which, once we consider \eqref{eqtwo}, becomes
\begin{equation}
\partial_t \Delta \beta = \left( {\partial \beta \over \partial \rho}\right)_{x_\mathrm{p}}  \partial_t \Delta \rho +  \left( {\partial \beta \over \partial x_\mathrm{p}}\right)_{\rho} {\gamma \over n} \Delta \beta  \ .
\end{equation}
That is, we have
\begin{equation}
\partial_t \Delta \beta -  \mathcal A  \Delta \beta = \mathcal B \partial_t \Delta \rho \ , 
\label{eqfour}
\end{equation}
with
\begin{equation}
\mathcal A =  \left( {\partial \beta \over \partial x_\mathrm{p}}\right)_{\rho} {\gamma \over n} \ , \qquad 
\mathcal B = \left( {\partial \beta \over \partial \rho}\right)_{x_\mathrm{p}} \ .
\end{equation}

The coefficients $\mathcal A$ and $\mathcal B$ are time independent, as they are evaluated for the equilibrium background, so if we work in the frequency domain (as we typically do when we consider stellar oscillations) then we have a harmonic time dependence $e^{i\omega t}$ and it follows that
\begin{equation}
\Delta \beta = {\mathcal B \over 1 + i \mathcal A/\omega} \Delta \rho \ .
\label{dbeta1}
\end{equation}
This relation is commonly taken as the starting point for discussions of bulk viscosity, see \cite{2018ASSL..457..455S} for a recent review of this issue. Here we want to make a slightly different emphasis.

Let us consider the timescales involved. Noting that $\mathcal A$ needs to be negative in order for the system to relax towards equilibrium, we introduce  a characteristic reaction time as 
\begin{equation}
t_R = -{1 \over \mathcal A}  \ .
\end{equation}
Then we see that, if the reactions are fast compared to the dynamics (associated with a timescale $\sim 1/\omega$) then $|t_R \omega| \ll 1$ and we have $\Delta \beta \approx 0$.
In effect, the fluid elements reach equilibrium before executing an oscillation. The fluid remains in beta-equilibrium and hence the perturbations are (effectively) barotropic. 

As a ballpark estimate of the relevant timescale, we draw on the discussion by \cite{2002A&A...394..213H} and assume
\begin{equation}
 t_R \sim 10^{13} \left( {10^8 \mathrm{K} \over T}\right)^6\ \mathrm{s}   
\end{equation}
for the modified Urca reactions (ignoring density dependence as we only need a rough idea here). This estimate suggests that---for all modes/rotating rates we may conceivably be interested in---mature neutron star matter will not be in the fast-reaction regime. We have to consider the slow-reaction (stratified) problem. 

In the limit of slow reactions we have $|t_R \omega| \gg 1$ and we can Taylor expand \eqref{dbeta1} to get
\begin{equation}
\Delta \beta \approx \mathcal B \left(  1-  i \mathcal A/\omega \right)  \Delta \rho \approx \mathcal B \Delta \rho \ .
\label{dbeta}
\end{equation}

Using this result in 
\begin{equation}
\Delta p = \left( {\partial p \over \partial \rho} \right)_{\beta} \Delta \rho +  \left( {\partial p \over \partial \beta} \right)_{\rho} \Delta \beta  \ ,
\label{eqone}
\end{equation}
we have
\begin{equation}
\Delta p = \left[ \left( {\partial p \over \partial \rho} \right)_{\beta} + \left( {\partial p \over \partial \beta} \right)_{\rho} \left( {\partial \beta \over \partial \rho}\right)_{x_\mathrm{p}} \right]  \Delta \rho  \ .
\label{eqfive}
\end{equation}
Comparing to \eqref{Gamma1}, we have an expression for $\Gamma_1$ in terms of thermodynamical derivatives:
\begin{equation}
    {p\Gamma_1\over \rho} = \left( {\partial p \over \partial \rho} \right)_{\beta} + \left( {\partial p \over \partial \beta} \right)_{\rho} \left( {\partial \beta \over \partial \rho}\right)_{x_\mathrm{p}} = \left( {\partial p \over \partial \rho } \right)_{x_\mathrm{p}} \ ,
\end{equation}
where the last equality---demonstrated by \cite{2019MNRAS.489.4043A}---holds as this limit represents frozen composition ($\Delta x_\mathrm{p}=0$). In this case we need to consider the impact of composition stratification on the fluid dynamics.

Moving on, introducing the gravitational acceleration (for a spherical star)
\begin{equation}
    g = - {1\over \rho} {dp \over dr} = {d \Phi \over dr} \ ,
    \end{equation}
we see that \eqref{Gamma1} leads to 
\begin{equation}
    \delta p = {p\Gamma_1 \over \rho} \delta \rho + \rho g \xi^r \left( 1 - {\Gamma_1 \over \Gamma} \right)  \ .
\end{equation}
It is important to note that the composition of matter impacts on both terms on the right-hand side of this relation. 

In the following, when we consider the impact of composition stratification on the oscillations of a slowly rotating star, it is convenient to consider the Brunt-V\"ais\"al\"a frequency, given by 
\begin{equation}
\mathcal N^2 = { \rho g^2 \over p}  \left( {1\over \Gamma} - {1\over \Gamma_1} \right) \ .
\end{equation}
This relation illustrates some of the subtleties we need to consider. For example,  if $\Gamma$ and $\Gamma_1$ are taken to be constant (as is commonly assumed) then $\mathcal N^2$ must diverge as we approach the surface of the star, where $p/\rho\to 0$. The only way to avoid this problem is if  the fluid becomes barotropic in the low-density limit, as we then have $\Gamma_1\to\Gamma$ as $\rho \to 0$. Also, it is important to keep in mind that the assumption that $\Gamma_1$ is constant is different from holding $\mathcal N^2$ fixed. However, as we will soon see, neither assumption is realistic.

The equation of state relation allows us to remove the density perturbation from the discussion. We then have 
\begin{equation}
 \delta \rho = {\rho \over p \Gamma_1} \delta p - { \rho g\over c_s^2}\xi^r \left(  {\Gamma \over \Gamma_1} -1 \right)  = {1\over c_s^2} \delta p - {\mathcal N^2 \over g^2} \left( \delta p - \rho g \xi^r \right) = {1\over c_s^2} \delta p - {\mathcal N^2 \over g^2} \Delta p \ .
    \label{drhoeq}
\end{equation}
Evidently, and quite intuitively, stratification does not affect incompressible flows, for which $\Delta \rho = 0 \Longrightarrow \Delta p = 0$.

For later convenience, if we introduce the density scale height
\begin{equation}
    {1\over H} = {1\over \rho } {d\rho \over dr}
\Longrightarrow 
    H = - {c_s^2 \over g} \ ,
\end{equation}
we may remove one of the three background quantities from the discussion. 

\subsection{The impact of stratification}

\begin{figure}
\begin{center}
\includegraphics[width=0.45\textwidth]{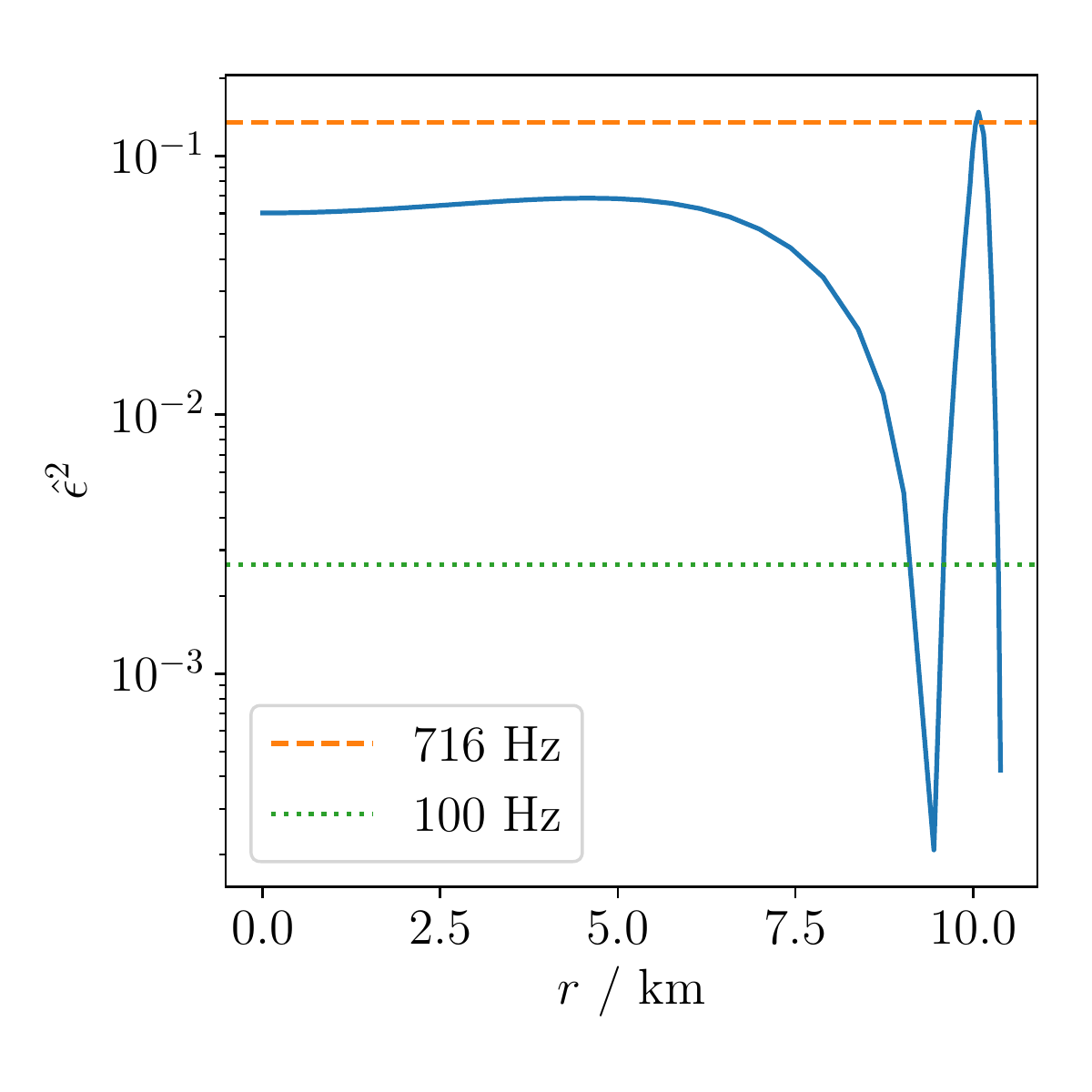}%
    \includegraphics[width=0.45\textwidth]{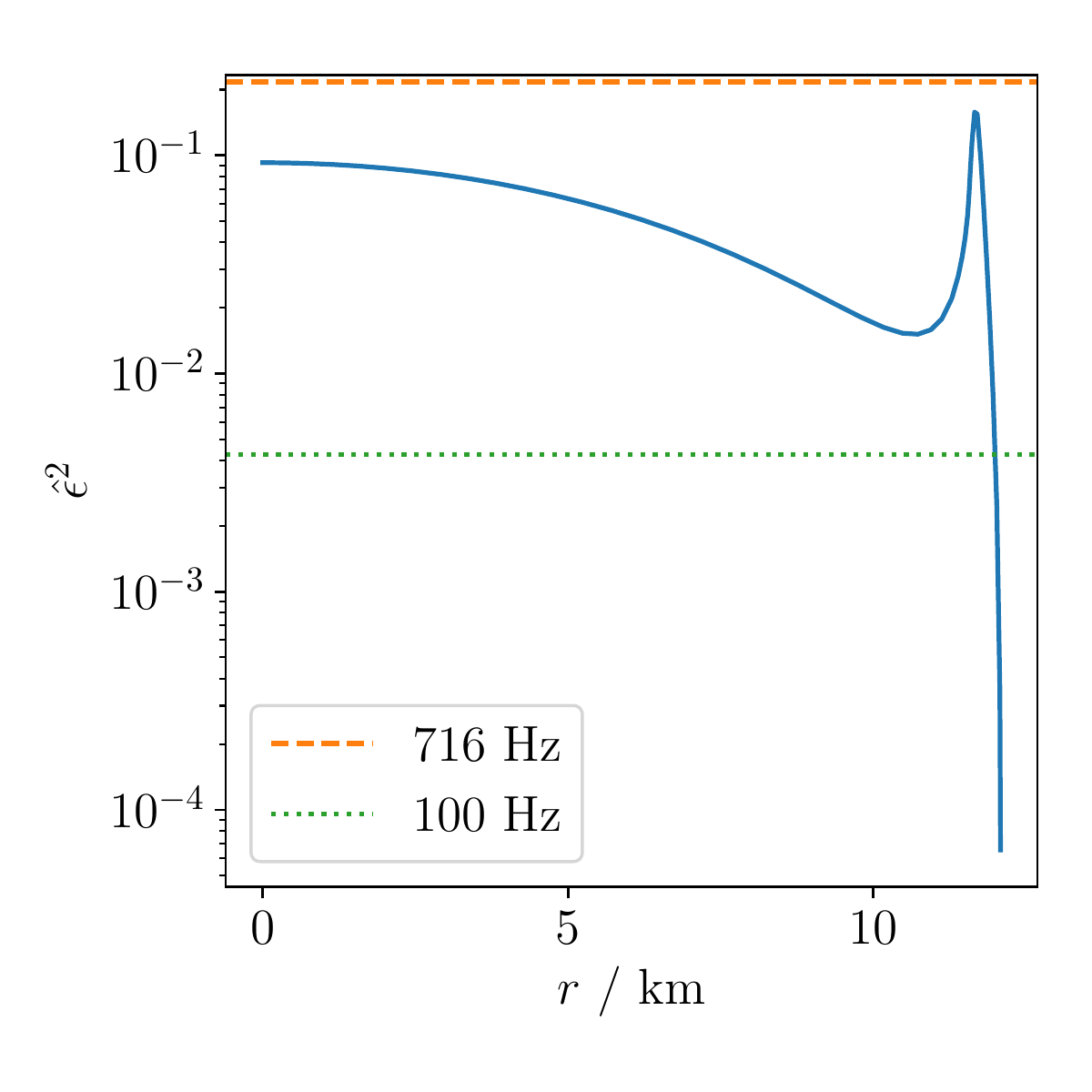}
\end{center}
   \caption{
    The dimensionless quantity $\hat{\epsilon}^2$ inside two $M = 1.4 \, M_\odot$ neutron stars described by the BSk19 (left panel) and BSk21 (right panel) equations of state from \citet{2013A&A...559A.128F}. The two models are suitably indicative because the dependence of the proton fraction with density is very different in the two cases (see Figure~4 from \citet{2013A&A...560A..48P}). The stellar models are obtained by integrating the relativistic equations of stellar structure with appropriate generalisations for the Brunt-V{\"a}is{\"a}l{\"a} frequency $\mathcal{N}$ and gravitational acceleration $g$. The horizontal lines indicate the value of $\epsilon^2=\Omega^2/\Omega_0^2$ for the two spin frequencies 100 Hz and 716 Hz, the latter representing the fastest known spinning neutron star \citep{2006Sci...311.1901H}.}
        \label{epshat}
\end{figure}

As already stated,  we will work with the displacement rather than the velocity perturbations in the following. We may then write 
 the perturbed continuity equation
\begin{equation}
 \delta \rho + \nabla_i (\rho \xi^i)=0 \ ,
\end{equation}
as (ignoring the rotational deformation of the background star, as advertised, 
and making use of the  equation of state relation \eqref{drhoeq})
\begin{equation}
     \nabla_i \xi^i=  - {1\over \rho c_s^2} \left(1   - {{\mathcal N}^2 c_s^2 \over g^2} \right)  \delta p + {g\over c_s^2} \left( 1 - { \mathcal N^2 c_s^2 \over g^2}  \right) \xi^r
     =- {1\over \rho c_s^2} \left(1   - {{\mathcal N}^2 c_s^2 \over g^2} \right)  \Delta p \ .
\end{equation}
 Clearly, it makes sense to consider the dimensionless quantity 
 \begin{equation}
     \hat \epsilon^2 = {{\mathcal N}^2 c_s^2 \over g^2} \ .
 \end{equation}
An example of this quantity, for the BSk19 and BSk21 equations of state \citep{2013A&A...559A.128F,2013A&A...560A..48P} (two models with very different  proton fraction profiles), is provided in Figure~\ref{epshat}. The results show that  $\hat \epsilon^2$ typically varies by about an order of magnitude throughout the star's core, reaches a peak in the low-density region and then drops sharply towards the surface. This behaviour will guide the discussion in the following.

Next consider the Euler equations (in a rotating frame), which take the form
\begin{equation}
-\omega^2\xi_i + 2i\omega\epsilon_{ijk} \Omega^j \xi^k + {1\over \rho} \nabla_i \delta p - {1\over \rho^2} \delta \rho \nabla_i p + \nabla_i \delta \Phi = 0  \ .
\label{xieul}
\end{equation}
For an axisymmetric rotating star, the $\varphi$-component becomes (noting that all perturbations behave as $e^{im\varphi}$)
\begin{equation}
    im  \left( \frac{\delta p}{\rho} + \delta \Phi \right)  = \omega^2 \xi_\varphi - 2i\omega \epsilon_{\varphi jk} \Omega^j \xi^k  \ .
\end{equation}
In essence, given that the inertial modes we are interested in have frequency $\omega \sim \Omega$, we must have $\delta p \sim \mathcal O(\Omega^2)$ in order to have $\xi^i \sim \mathcal O(1)$. The radial component of the Euler equation then tells us that we must also have $\delta \rho \sim \mathcal O(\Omega^2)$ and the equation of state relation \eqref{drhoeq} then leads to 
\begin{equation}
    {\mathcal N^2 \over g} \xi^r \lesssim \mathcal O(\Omega^2) \ .
\end{equation}
For a given stratification, expressed in terms of $\mathcal N^2$, this constrains the slow-rotation ordering of the radial displacement. This accords with the point we made earlier \citep{1999ApJ...521..764L}. 

It is, however,  useful to make this argument more precise. For inertial modes of uniformly rotating stars it is natural to use the slow-rotation expansion with $\epsilon = {\Omega/\Omega_0}$
as the small parameter, noting that the Kepler break-up frequency corresponds to roughly 
\begin{equation}
\Omega_K \approx \frac{2}{3} \sqrt{\pi G \rho_0} \Longrightarrow \Omega_K^2 \approx  {1\over 3} \Omega_0^2 \Longrightarrow \epsilon_K \approx 0.6 \ .
\end{equation}
 Given the previous discussion we know that we should have $\delta p = \epsilon^2 \delta \tilde p$
using the tilde to indicate an $\mathcal O(1)$ quantity (notation we will adopt in the following), and $\delta \rho = \epsilon^2 \delta \tilde \rho$, as well. 
If, in addition, we focus on inertial modes we know that $\omega\sim \Omega$ and it is evident from \eqref{xieul} that we do not have to consider terms of order $\epsilon$; we can take $\epsilon^2$ as the small rotation parameter\footnote{Note that this is not true for modes that have a finite frequency in the non-rotating limit.}. Hence, we expand the mode frequency as 
\begin{equation}
    \omega_n = \Omega \left( \omega_0 + \epsilon^2 \tilde \omega_2 \right) \ .
    \label{omex}
\end{equation}

The equation of state relation \eqref{drhoeq} then becomes
\begin{equation}
\epsilon^2 c_s^2 \delta \tilde \rho = \epsilon^2  \delta \tilde  p - \hat \epsilon^2 \left( \epsilon^2 \delta \tilde p - \rho g \xi^r\right) \ .
\end{equation}
Let us see what we learn from this.

Noting that the  Kepler frequency corresponds to $\epsilon \lesssim 1$ while the results from Figure~\ref{epshat} show that $\hat \epsilon^2< 1$ as well, we may consider a double expansion in $\epsilon$ and $\hat \epsilon$. Of course, since $\hat \epsilon$ varies throughout the star this would have to be a local argument. The first few terms of such an expansion would  be
\begin{equation}
\xi^i \approx \xi_0^r + \hat \epsilon^2 \hat \xi_2^r + \epsilon^2 \xi_2^r  
\end{equation}
and
it is easy to see that we need to consider several different cases. 

First, suppose $\hat \epsilon \gg \epsilon$ as would be the case if we combine the results in Figure~\ref{epshat} with a fairly slowly spinning star. Effectively, this situation corresponds to taking $\hat \epsilon \sim \mathcal O(\epsilon^{0})$ and---formally letting $\xi_0^r + \hat \epsilon^2 \hat \xi_2^r\Longrightarrow \xi_0^r$---it follows that
\begin{eqnarray}
\xi_0^r  &=&  0
 \ , \quad\mbox{at order }\ \epsilon^{0} \ , \label{Wstrat} \\
c_s^2 \delta \tilde \rho  &=&  \delta \tilde p - \hat \epsilon^2( \delta \tilde p - \rho g \xi^r_2)
 \ , \quad \mbox{at order }\ \epsilon^2 \ .
\end{eqnarray}
This is the usual stratified r-mode ordering (see, for example, \cite{1981A&A....94..126P}), but it is not clear that this is the case we should consider. The argument would only apply to slow and perhaps moderately fast spinning neutron stars (say, the 100~Hz case illustrated in Figure~\ref{epshat}), but not the fastest observed systems.

Instead, the results in Figure~\ref{epshat} suggest that, for stars spinning with $\epsilon \sim 0.3$ (about half the Kepler rate), roughly corresponding to the fastest known pulsars with spin frequency close to 700~Hz,  we should instead consider $\hat \epsilon \sim \mathcal O(\epsilon)$, in which case we may ignore terms of order $\epsilon^2 \hat \epsilon^2$ in the expansion. This then leads to
\begin{equation}
c_s^2 \delta \tilde \rho  =  \delta \tilde p + \hat \epsilon^2 \rho g \xi^r_0
\ , \quad\mbox{at order }\ \epsilon^2 \ .
 \label{neweps}
\end{equation}
This is different from what we usually assume, yet seems a  case we ought to consider. Perhaps significantly, the relation suggests that we may have $\xi_0^r \neq 0$, which changes the mode structure. The model is also interesting as it limits to the barotropic case in regions where $\hat \epsilon^2\to 0$.

Finally, from Figure~\ref{epshat} it is clear that there will always be always a  low-density region where
 $\hat \epsilon \lesssim \mathcal O(\epsilon^{2})$. If the  $\hat \epsilon^2$ are smaller than (say) $\epsilon^4$ then we may ignore the stratification and consider the barotropic result
\begin{eqnarray}
c_s^2 \delta \tilde \rho  &=  \delta \tilde p 
 \ , \quad &\mbox{at order }\ \epsilon^2 \ .
\end{eqnarray}
If these assumptions hold throughout the star, then we must end up with the inertial modes from \cite{1999ApJ...521..764L}. Globally, this is unlikely to be the relevant case but Figure~\ref{epshat} suggests that we always have to to consider the region close to the surface as barotropic. This impacts on the surface boundary condition and may, in turn, also affect the modes.

Clearly, the profile for $\hat \epsilon$ is fixed for any given stellar model, while $\epsilon$ can be varied (up to the Kepler limit, which means that we should have  $\epsilon \lesssim 0.6$). From Figure~\ref{epshat} it is easy to see that all the suggested orderings  may apply locally in a neutron star core, making the formulation of  consistent model tricky. The main conclusion is that we have to consider all results obtained with the standard ``constant-$\Gamma_1$'' prescription as unrealistic. In fact, if we take the results in Figure~\ref{epshat} at face value---and there is no reason why we should not---then we have to reconsider our strategy for the fastest spinning stratified neutron stars. For the core of these stars,  relation \eqref{neweps} should apply, in which case the perturbation problem is closer to---but not exactly the same as---that for inertial modes \citep{1999ApJ...521..764L}. To what extent this affects the mode frequencies remains to be established.

\section{Formulating the mode problem}

A mode solution to the perturbation problem must satisfy the perturbation equations (obviously) and relevant boundary conditions (typically, regularity at the centre of the star and the vanishing of the Lagrangian variation of the pressure at the star's surface).  It is well established that the equations allow for oscillation modes of  different character. Moreover, the problem gets richer as more detailed neutron star physics is considered. Somewhat simplistically, each aspect of physics added to the model---matter composition, rotation, superfluidity, electromagnetism, elasticity...---brings new families of modes into play. This makes the general problem complex. 

In order to build useful intuition, it is natural to focus on particular aspects. If we are mainly interested in the oscillations of a rotating star,  the natural starting point would be to work out how the rotation impacts on modes that exist already in a non-rotating star.  For a mode with frequency $\omega_n$ in the non-rotating star, the strategy is---at least in principle---fairly straightforward, although the mode calculation may get quite involved as the rotation couples different multipole contributions. Still, for slow to moderate rotation rates, this problem can be dealt with perturbatively as long as $\Omega \lesssim \omega_n$. This should be the case for the fundamental mode of the star, which has frequency of order the Kepler breakup frequency, and the (even higher frequency) pressure p-modes. We will not consider those problems here. We will also not consider the rotational corrections to the gravity g-modes, a slightly more subtle issue given that the high-overtone g-modes are expected to have very low frequencies in a non-rotating star. Hence, for these modes the Coriolis force may dominate over the buoyancy already at fairly low rotation rates and they would then become part of the problem we are considering. 
An example of this behaviour can be found in the study by \cite{2000ApJ...529..997Y}, where it is shown how inertial modes are strongly modified when the buoyancy force becomes comparable to, or stronger
than the Coriolis force.

In this exploratory analysis our main focus is on the qualitative nature of the low-frequency modes of a rotating star.  In this spirit, we rely on simpliying assumptions. In particular,  following \citet{1997ApJ...491..839L}, we  ignore the change in shape of the background star associated with the centrifugal force. This assumption is not expected to affect the qualitative nature of the problem.

As we have already seen, the  discussion necessarily gets somewhat involved and some of the issues we need to consider are subtle. In general, we need to consider both the impact of rotation on oscillation modes that exist already in a non-rotating star and modes that are brought into existence when we consider the impact of the Coriolis force. Given this, it makes sense to work out the general perturbation equations to first slow-rotation order. This involves making choices already at the outset.

There are three common strategies for investigating the oscillations of rotating stars \citep{1989nos..book.....U}. The first, and formally most elegant, approach expresses the rotational corrections to a given mode as a sum over all the modes of the corresponding non-rotating star (which form a suitable complete basis as long as we ignore dissipation). The second option builds on an explicit expansion in angular harmonics while the third involves time evolving the perturbation equations. The last strategy has the advantage that one can readily deal with fast spinning stars, for which the algebra of the other approaches becomes daunting, but it also has the drawback that one loses track of the fine details of the problem (you get what you get from the simulation, depending on the chosen initial data). Examples of work in this direction can be found in \citet{10.1046/j.1365-8711.2002.05566.x,10.1111/j.1365-2966.2009.14408.x} for  Newtonian models and \citet{PhysRevD.80.064026,PhysRevLett.107.101102,PhysRevD.83.064031,2021FrASS...8..166K} for efforts in relativity. 
In the following we will carry out an expansion in harmonics. This approach has the advantage that it highlights the nature of the fluid motion.

Assuming that the oscillation modes---with label $n$ and frequency $\omega_n$---are associated with a displacement vector 
\begin{equation}
    \xi^i_n(t,r,\theta,\varphi) = \hat \xi^i (r,\theta,\varphi) e^{i\omega_n t} \ ,
\end{equation}
(where the hat indicates a quantity that is independent of $t$ and we adopt the \cite{1999ApJ...521..764L} sign convention)
we  have  (in a coordinate basis with $e_r^i = \partial \boldsymbol r/\partial r$ etc)
\begin{equation}
 \hat \xi^i = \sum_l \left[ { 1 \over r} W_l Y_l^m {e}^i_r + \left({ 1 \over r^2}  V_l \partial_\theta Y_l^m + { m \over r^2 \sin\theta} U_l Y_l^m
 \right){e}_\theta^i + { i \over r^2 \sin^2 \theta}  \left( m V_l Y_l^m + U_l \sin \theta \partial_\theta Y_l^m  \right) {e}_\varphi^i \right] \ ,
 \label{multipolesum}
\end{equation}
where we refer to $W_l$ and $V_l$ as polar perturbations, while $U_l$ is axial (and noting that the $m$-multipoles decouple for an axisymmetric system, like a rotating star). For a given multipole $l$, these perturbations have different parity. This follows since the equilibrium state of a rotating star is invariant under the parity transformation (defined by a reflection through the origin,
$\theta \to \pi - \theta$ and $\varphi\to \varphi+\pi$), the linear perturbations have definite parity for this transformation. Alternatively, the different modes are sometimes described as even and odd, see for example, \citet{1987MNRAS.224..513L}. 

Along with the decomposition of the displacement, all scalar perturbations are expanded in spherical harmonics. That is,  we have 
\begin{equation}
    \delta \rho_n = \delta \hat \rho (r,\theta,\varphi)  e^{i\omega_n t} \ ,
\end{equation}
with (dropping the hats on the individual $l$-multipole components to keep the equations that follow as tidy as possible)
\begin{equation}
\delta \hat \rho = \sum_l \delta \rho_l Y_l^m 
\end{equation}
and similar for all other scalar quantities.
We also know that the rotating equilibrium remains spherical to linear order in $\Omega$ so all background quantities depend only on $r$ as long as we consider the first order slow-rotation corrections. Working to this order of approximation, let us summarise the equations we need.

\subsection{The perturbation equations}

First, it follows
from \eqref{multipolesum}  that 
\begin{equation}
  \nabla_i \hat \xi^i =   \sum_l {1\over r^2} \left[  \partial_r \left(r W_l \right) - l(l+1) V_l\right] Y_l^m \ .
\end{equation}
This result is important because it shows that---up to order $\Omega$---only the polar contributions to a given mode $[W_l,V_l]$ contribute to the density perturbation $\delta \rho_l$.  This is brought out by the continuity equation, which leads to (changing $l\to j$ for consistency with the recurrence relations to be derived in the following)
\begin{equation}
     {1 \over r^2} \partial_r \left( r\rho W_j\right)   - j(j+1) {\rho \over r^2}  V_j =   -\delta \rho_j \ .
         \label{conteq}
\end{equation}
Notably---as long as we ignore the rotational deformation---this equation does not involve the coupling of different multipoles.

Turning to the perturbed Euler equations, in the rotating frame we have 
\begin{equation}
-\omega_n^2\hat \xi_i + 2 i\omega_n  \epsilon_{ijk}\Omega^{j} \hat \xi^k + { 1 \over \rho} \nabla_{i}\delta \hat p - { 1 \over \rho^2} \delta \hat \rho \nabla_{i} p + \nabla_i \delta \tilde \Phi
 = 0 \ ,
\end{equation}
leading to the radial component 
\begin{equation}
\mathcal E_r = {1\over r} \sum_l \left[ \left( - \omega_n^2 W_l  +   2  m \omega_n \Omega V_l
+ r\partial_r \delta  \Phi_l + { 1 \over \rho} r\partial_r \delta  p_l - { 1 \over \rho^2} \delta  \rho_l r\partial_r  p\right)  Y_l^m + 2 \omega_n \Omega U_l \sin \theta \partial_\theta Y_l^m  \right] = 0  \ ,
\end{equation}
the $\theta$ component
\begin{multline}
\mathcal E_\theta =   {1\over \sin\theta} \sum_l \Bigg[ - \omega_n^2 \left(  V_l \sin\theta \partial_\theta Y_l^m + m U_l Y_l^m
 \right) + 2\omega_n \Omega   \cos\theta  \left( m V_l Y_l^m + U_l \sin \theta \partial_\theta Y_l^m  \right) \\
 + \left(  \delta  \Phi_l  + \frac{\delta  p_l }{\rho} \right) \sin \theta \partial_\theta Y_l^m\Bigg]= 0  
 \label{theteul}
\end{multline}
and the $\varphi$ component
\begin{multline}
\mathcal E_\varphi =  i \sum_l \Bigg\{ -  \omega_n^2 \left( m V_l Y_l^m + U_l \sin \theta \partial_\theta Y_l^m  \right) \\ + 2 \omega_n \Omega \sin\theta  \left[ \sin\theta W_l Y_l^m +  \cos\theta \left(  V_l \partial_\theta Y_l^m + {m \over \sin\theta} U_l Y_l^m
 \right)  \right] + m  \left(  \delta  \Phi_l  + \frac{\delta  p_l }{\rho} \right) Y_l^m \Bigg\}
 =0 \ .
 \label{phieul}
\end{multline}

One possible strategy would be to work with these equations as they are, deal with the fact that different multipoles couple head on and solve the problem numerically (see, for example, \citet{2006MNRAS.365..677L}). However, this may not be the most ``transparent'' option as it obscures the nature of the different mode solutions. As we will see later, it follows from the angular components \eqref{theteul} and \eqref{phieul} that we have to consider coupling between the $l$-components and the ones for $l\pm2$. This leads to a larger set of equations to solve so it makes to ask if this coupling can be avoided. It turns out that it cannot, but we can find a somewhat more ``intuitive'' set of equations to solve. 

From the Euler equations it is easy to see that it makes sense to introduce
\begin{equation}
    \delta \mathcal U_l = {\delta p_l \over \rho} +  \delta \Phi_l \ ,
\end{equation}
(not to be confused with the axial amplitude $U_l$).
This variable is used both in the  classic (dimensionless) formulation from \citet{1989nos..book.....U} and  the two-potential formalism used by, for example, \citet{1999PhRvD..59d4009L}.

 Making use of the standard recurrence relation
\begin{equation}
\sin \theta \partial_\theta Y_l^m = l \mathcal Q_{l+1} Y_{l+1}^m - (l+1) \mathcal Q_l Y_{l-1}^m \ ,
\label{sinth}
\end{equation}
where
\begin{equation}
\mathcal Q_l = \left[ {(l-m)(l+m) \over (2l-1) (2l+1)}\right]^{1/2} \ ,
\label{Qdef}
\end{equation}
the radial Euler equations leads to
\begin{multline}
\sum_l \Bigg\{ \left[ - \omega_n^2 W_l +   2  m \omega_n \Omega V_l
+ r\partial_r \delta  \mathcal U_l  + { r \over \rho^2} \left( \delta p_l \partial_r \rho - \delta  \rho_l r \partial_r  p\right) \right] Y_l^m \\
+ 2 \omega_n \Omega U_l \left( l \mathcal Q_{l+1} Y_{l+1}^m - (l+1) \mathcal Q_l Y_{l-1}^m  \right) \Bigg\} = 0 \ .
\end{multline}
Multiplying this by $Y_j^m$,  integrating over the angles and executing the sum over $l$ we have the recurrence relation
\begin{equation}
 - \omega_n^2 W_j +   2  m \omega_n \Omega V_j
+ r\partial_r \delta  \mathcal U_j  + { r \over \rho^2} \left( \delta p_j \partial_r \rho - \delta  \rho_j r \partial_r  p\right)
+ 2 \omega_n \Omega \left[  (j-1) \mathcal Q_j U_{j-1}  - (j+2) \mathcal Q_{j+1} U_{j+1} \right] = 0\ .
\label{radeq}
\end{equation}

Next we consider the combination (the radial component of the vorticity equation)
\begin{multline}
   \partial_\varphi \mathcal E_\theta - \partial_\theta \mathcal E_\varphi 
    =  -i \omega_n \sin \theta \sum_l \Bigg\{ \left[ l(l+1) \omega_n  - 2m\Omega \right] U_l Y_l^m - 2\Omega \left[   l(l+1) \cos\theta Y_l^m  + \sin \theta \partial_\theta Y_l^m \right] V_l \\
  + 2\Omega  \left[ 2 \cos \theta  Y_l^m + \sin \theta \partial_\theta Y_l^m \right] W_l
 \Bigg\} = 0  \ ,
 \label{radvor}
\end{multline}
  where we have made use of Legendre's equation
  \begin{equation}
      \partial_\theta \left( \sin \theta \partial_\theta Y_l^m \right) = \left[ {m^2 \over \sin \theta} - l(l+1) \sin \theta \right] Y_l^m \ ,
  \end{equation}
  to simplify the result.

Using the previous recurrence relation \eqref{sinth}, along with 
\begin{equation}
\cos \theta Y_l^m = \mathcal Q_{l+1} Y_{l+1}^m + \mathcal Q_l Y_{l-1}^m \ ,
\label{costh}
\end{equation}
we arrive at
the recurrence relation
\begin{equation}
 \left[ j(j+1) \omega_n  - 2m\Omega \right] U_j + 2\Omega (j+1) [ W_{j-1} - (j-1) V_{j-1}]  \mathcal Q_{j}  - 2\Omega j[ W_{j+1} + (j+2) V_{j+1}] \mathcal Q_{j+1} = 0    \ .
 \label{axrec}
\end{equation}

Keeping Legendre's equation in mind, it may also be useful to consider the combination \citep{1996A&A...311..155L,2006PhRvD..74d4040G}
\begin{multline}
    \partial_\theta ( \sin \theta \mathcal E_\theta) + {1\over \sin \theta} \partial_\varphi \mathcal E_\varphi 
 =  \sum_l \left\{  \omega_n \left[ l(l+1) \omega_n  -2m\Omega\right] V_l -2m \omega_n \Omega W_l  - l(l+1)  \left(  \delta  \Phi_l  + \frac{\delta  p_l }{\rho} \right) \right\}\sin \theta Y_l^m \\
 - \sum_l 2\omega_n \Omega \sin\theta \left[  l(l+1) \cos\theta Y_l^m  + \sin \theta \partial_\theta Y_l^m  \right]  U_l = 0 \ ,
\end{multline}
which leads to 
\begin{multline}
 \omega_n \left[ j(j+1) \omega_n  -2m\Omega\right] V_j -2m \omega_n \Omega W_j  - j(j+1)  \delta  \mathcal U_j   \\
 -  2\omega_n \Omega \left[  (j-1)(j+1) \mathcal Q_{j} U_{j-1} +  j (j+2) Q_{j+1}  U_{j+1} \right] = 0 \ .
 \label{diveq}
\end{multline}

For inertial modes, it is notable that the radial vorticity equation \eqref{axrec}, links only variables that have a leading order contribution. Another such relation follows from the $\theta$ component of the vorticity equation
\begin{multline}
\partial_r \mathcal E_\varphi - \partial_\varphi \mathcal E_r =
    \omega_n \left[2 \Omega \left(1 - \mathcal Q_j^2 - \mathcal Q_{j+1}^2 \right) r\partial_r W_j + m\omega_n W_j \right] 
    \\
     -\omega_n \left[  \left\{ m\omega_n +2\Omega\left[ (j+1) \mathcal Q_j^2- j\mathcal Q_{j+1}^2 \right]  \right\}  r\partial_r V_j + 2m^2\Omega V_j \right] \\
    - \omega_n \left[ \left( \omega_n (j-1)  - 2m\Omega \right) r\partial_r U_{j-1}  +2m (j-1) \Omega  U_{j-1} \right] \mathcal Q_{j}   \\
+\omega_n  \left\{ \left[ \omega_n (j+2)  + 2m\Omega \right] r\partial_r U_{j+1}  + 2m (j+2) \Omega  U_{j+1} \right\} \mathcal Q_{j+1}\\
+ 2\omega_n \Omega \left[ (j-2) r\partial_r V_{j-2} - r \partial_r W_{j-2} \right] \mathcal Q_{j-1} \mathcal Q_{j}  - 2\omega_n \Omega \left[ (j+3) r\partial_r V_{j+2} + r \partial_r W_{j+2} \right] \mathcal Q_{j+2} \mathcal Q_{j+1}  \\ = -  {mr\over \rho^2} \left(\delta \rho_j \partial_r p - \delta p_j \partial_r \rho \right) 
\ .    \label{angvor}
\end{multline}
This equation is identical to equation (39) from Lockitch and Friedman, apart from the right-hand side which only vanishes for barotropic stars. In general, the equation is a bit messy as it couples radial derivatives of all displacement components and it also involves the $l\pm 2$ multipoles. However, we will find the equation useful for non-barotropic stars satisfying the traditional r-mode slow-rotation ordering as it simplifies considerably in that case. Finally, we also need the perturbed Poisson equation for the gravitational potential. However, we will make the Cowling approximation in our explicit examples (set $\delta \Phi_j=0$) so will not give the equation here. 

\section{The ``traditional'' r-modes}

As already advertised, we will focus on low-frequency modes such that $\Omega\gtrsim \omega_0$. This  includes high-order gravity g-modes and inertial modes. We have already seen that the perturbed Euler equations then imply that we must  have $[\delta \rho_l, \delta p_l, \delta \Phi_l] \sim \mathcal O(\Omega^2)$.
Moreover, the continuity equation requires the polar components $W_l$ and $V_l$ to be of the same order, and we already know that if we consider a strongly stratified star (with $\hat \epsilon \gg \epsilon$) then we must have $[W_l, V_l]\sim \mathcal O(\Omega^2)$.
We are then left to consider if it is possible to combine these assumptions with  $U_l \sim \mathcal O(1)$.
The answer to this question is affirmative, but it follows from \eqref{axrec} that we must then have
\begin{equation}
 \sum_l \left[ l(l+1) \omega_n  - 2m\Omega \right] U_l Y_l^m  = 0  \ . 
 \label{radvor3}
\end{equation}
That is, for given values of $l=l'$ (say) and $m$,  we may have $U_{l'}\neq 0$ as long as the leading order mode frequency is given by 
\begin{equation}
\omega_0 = {2m  \over l'(l'+1)} 
 \ . \label{rfreq}
\end{equation}
These are the r-modes \citep{1978MNRAS.182..423P, 1981A&A....94..126P,1982ApJ...256..717S}. In addition to the leading order displacement they will have polar components as well
as other axial multipoles $U_{l\neq l'}$, but  these enter at $\mathcal O(\Omega^2)$. We will deliberate on these contributions  in the following. 

With our conventions, the pattern speed of a mode is $- \omega / m$. Therefore, all r-modes travel in the same direction across the star (retrograde with respect spin, in the rotating frame).  We also note that there are no axisymmetric r-modes; we must have $m \neq 0$.

Focusing on the r-mode problem, we consider the perturbations for  mode solutions such that $\omega_n \sim \mathcal O (\Omega)$ and
\begin{equation}
    U_{l=l'}\sim \mathcal O(1) , \quad
      [W_l , V_l, \delta \rho_l, \delta p_l] \sim \mathcal O(\Omega^2) \ , \quad 
 U_{l}\sim \mathcal O(\Omega^2) \ \mbox{for}\ l\neq l' \ .
\end{equation}

As before, we use tildes to identify terms that enter at order $\Omega^2$, i.e. 
\begin{equation}
  \delta p_l = \epsilon^2 \delta \tilde p_l = \left( {\Omega\over \Omega_0} \right)^2 \delta \tilde p_l\ ,
\end{equation}
with  $\delta \tilde p_l\sim \mathcal O(1)$ by definition.
In addition, we need to keep track of the rotational correction to the frequency, so recall \eqref{omex} from which it is worth noting that assumed slow-rotation ordering is only valid as long as 
\begin{equation}
 {\tilde \omega_2 \over \omega_0} \epsilon^2 \ll 1 \ .
 \end{equation}

With these assumptions, the continuity equation \eqref{conteq} relates order $\Omega^2$ quantities, and we have
\begin{equation}
     {1\over r^2}   \partial_r \left(r \rho \tilde W_j \right) - {j(j+1) \rho \over r^2} \tilde V_j = - {1\over c_s^2} \delta\tilde  p_j + {\mathcal N^2 \over g^2} \left( \delta  \tilde p_j -  {\rho g \over r} \tilde W_j \right) \ .
    \label{continu1}
\end{equation}
So far, the different relations only involve single-multipole polar components. This changes when we turn to the perturbed Euler equations. 

Let us first consider the radial component of the vorticity equation \eqref{axrec}. We know already that, at leading order we may have a single axial contribution $U_{l'}\neq 0$ as long as the frequency is given by \eqref{rfreq}. 
However, at this point, 
we cannot determine the axial eigenfunction. Essentially, the leading order r-mode solutions are degenerate. To break this degeneracy, we need to go to higher orders, keeping in mind that the polar contributions enter at order $\Omega^2$. 

Equation \eqref{axrec} provides a recurrence relation involving these multipole contributions. At order $\Omega^3$ we have (with $j=l')$
\begin{equation}
 l'('l+1) \tilde \omega_2  U_{l'} + 2 (l'+1) [ \tilde W_{l'-1} - (l'-1) \tilde V_{l'-1}]  \mathcal Q_{l'}  - 2 l' [\tilde W_{l'+1} + (l'+2) \tilde V_{l'+1}] \mathcal Q_{l'+1} = 0   \ .
 \label{jl}
\end{equation}
This is the only relation we get that involves the leading order eigenfunction and the frequency correction $\tilde \omega_2$. However, for $j=l'+2$ we get from \eqref{axrec}
\begin{multline}
 \left[ (l'+2)(l'+3) \omega_0  - 2m \right] \tilde U_{l'+2} + 2 (l'+3) [ \tilde W_{l'+1} - (l'+1) \tilde V_{l'+1}]  \mathcal Q_{l'+2}  \\
 - 2 (l'+2)[ \tilde W_{l'+3} + (l'+4) \tilde V_{l'+3}] \mathcal Q_{l'+3} = 0  \ ,
 \label{jlp2}
\end{multline}
while $j=l'-2$ leads to
\begin{equation}
 \left[ (l'-2)(l'-1) \omega_0  - 2m \right] \tilde U_{l'-2} + 2 (l'-1) [ \tilde W_{l'-3} - (l'-3) \tilde V_{l'-3}]  \mathcal Q_{l'-2}  
 - 2 (l'-2) [ \tilde W_{l'-1} + l' \tilde V_{l'-1}] \mathcal Q_{l'-1} = 0 \ .
 \label{jlm2}
\end{equation}
 From these relations, the pattern is clear. At order $\Omega^2$, the r-mode solution  involves a number of multipoles. The question then becomes, does this sequence truncate?
To answer this question, first note that $\mathcal Q_m=0$, which helps establish the  lowest order term in the series. We have three options. First, we may have $l'=m$ in \eqref{jl}. In this case, the (leading order) $U_{l'=m}$ term  corresponds to the lowest order multipole in the solution. In the language of \citet{1999ApJ...521..764L}, the mode is axial-led. This case corresponds to the traditional $l'=m$ r-mode \citep{1978MNRAS.182..423P}. Another option would be to have $l'=m+2$ in \eqref{jlm2}. This would also lead to an axial-led mode, but now the lowest multipole is given by $\tilde U_{l'-2}\sim \mathcal O(\Omega^2)$.
A third option follows by setting $l'=m-1$ in \eqref{jlm2} which then decouples and from \eqref{jl} we arrive at a polar-led mode  with the lowest multipole contributions given by $[\tilde W_{l'-1},\tilde V_{l'-1}]$ (again at order $\Omega^2$). The main lesson here is that the nature of the r-modes is quite similar to that of the general inertial modes discussed by \citet{1999ApJ...521..764L}. Each mode has several multipole contributions, but the $U_{l'}$ term is elevated above the other contributions in the slow-rotation expansion. The  close relation between the two problems may not been very clearly explained in the existing literature. It is, however, important for what follows.

In order to complete the formulation of the problem, we will use the other vorticity equation \eqref{angvor}. With the 
 ordering we have, at order $\Omega^2$, this reduces to
\begin{multline}
     - \omega_0 \left\{[\omega_0 (j-1) -2m] r \partial_r U_{j-1} + 2m(j-1) U_{j-1} \right\} \mathcal Q_{j}  \\
    + \omega_0\left\{[\omega_0 (j+2)+2m]r\partial_r U_{j+1} + 2m(j+2))U_{j+1}\right\}  \mathcal Q_{j+1}    \\
    + {mr\over \rho g} {\mathcal N^2 \over \Omega_0^2} \left( \delta \tilde p_j -  {\rho g \over r} \tilde W_j \right) 
    =0 \ .
    \label{angrec}
\end{multline}

From this relation, we infer two relations involving the leading order $U_{l'}$ term. First, with $j=l'+1$ we have
\begin{equation}
      \left({2m\over l'+1} \right)^2 \mathcal Q_{l'+1} \left[  r \partial_r U_{l'} - (l'+1) U_{l'}\right]
    + {mr\over \rho g} {\mathcal N^2 \over \Omega_0^2} \left( \delta \tilde p_{l'+1} -  {\rho g \over r} \tilde W_{l'+1} \right) 
    =0 \ .
    \label{Uder1}
\end{equation}
Second, with $j=l'-1$ we get
\begin{equation}
     \left({2m\over l'} \right)^2 \mathcal Q_{l'}  \left[  r\partial_r U_{l'} + l' U_{l'} \right]
   + {mr\over \rho g} {\mathcal N^2 \over \Omega_0^2} \left( \delta \tilde p_{l'-1} -  {\rho g \over r} \tilde W_{l'-1} \right)
    =0 \ .
    \label{Uder2}
\end{equation}

Finally, we have the divergence equation \eqref{diveq}, which leads to
\begin{equation}
  - j(j+1)  \delta   \tilde {\mathcal U}_j 
 -  2\omega_0 \Omega_0^2  \left[  (j-1)(j+1) \mathcal Q_{j} U_{j-1} +  j (j+2) Q_{j+1}  U_{j+1} \right] = 0  \ .
\end{equation}
For $j=l'+1$ we have
\begin{equation}
    \delta   \tilde {\mathcal U}_{l'+1} = - {2\omega_0 l' \over (l'+1)} \Omega_0^2 \mathcal Q_{l'+1} U_{l'} =  - {4m \Omega_0^2  \over (l'+1)^2}  \mathcal Q_{l'+1} U_{l'}  \ ,
    \label{duplus}
\end{equation}
while $j=l'-1$ leads to
\begin{equation}
    \delta   \tilde {\mathcal U}_{l'-1} = - {2\omega_0 (l'+1) \over (l')} \Omega_0^2 \mathcal Q_{l'} U_{l'} =  - {4m \Omega_0^2  \over (l')^2}  \mathcal Q_{l'} U_{l'} \ .
    \label{duminus}
\end{equation}

In essence, if we want to determine the leading order eigenfunction and the frequency correction $\tilde \omega_2$, we need to solve a coupled system for $U_{l'}$ and $\tilde W_{l'\pm 1}$. The other contributions to the mode-solution (like $\tilde U_{l'\pm2}$)  can be calculated as a second step. 

Finally, the mode-solution must satisfy the condition that the Lagrangian perturbation in the pressure vanishes at the surface. That is, we require
\begin{equation}
\Delta \tilde p_l =\delta \tilde p_l - {\rho g \over r} \tilde W_l = 0 \quad \mbox{at} \quad r=R \ .
    \label{surfcon}
\end{equation}

At this point, we may return to the question of whether the multipole sum truncates for the r-modes. First, the relation \eqref{angrec} also tells us, for $j=l'+3$ and $j=l'-3$, respectively, that  we must have (as long as $\mathcal N^2\neq 0$)
\begin{equation}
    \delta \tilde p_{l'\pm3} - {\rho g\over r} \tilde W_{l'\pm 3} = \Delta \tilde p_{l'\pm 3} =  0 \ .
\end{equation}

Second, in \eqref{duplus} and \eqref{duminus}, we can use $j=l'\pm 2$ to show that we must have
\begin{equation}
     \delta\tilde {\mathcal U}_{l'\pm 3} = \delta \tilde \Phi_{l'\pm 3}+  {\delta \tilde p_{l'\pm 3} \over \rho }= 0 \ .
\end{equation}

Third, with the assumed r-mode ordering, the radial component of the Euler equations \eqref{radeq} leads to
\begin{multline}
 r\partial_r \delta \tilde{  \mathcal U}_j  + 2 \omega_0  \Omega_0^2 \left[  (j-1) \mathcal Q_j U_{j-1}  - (j+2) \mathcal Q_{j+1} U_{j+1} \right] 
 \\ = - { r \over \rho^2} \left( \delta \tilde p_j \partial_r \rho - \delta  \tilde \rho_j r \partial_r  p\right) =  { \mathcal N^2  r \over  g}  \left( {1\over \rho} \delta \tilde p_j - { g \over r} \tilde W_j \right)
 =  { \mathcal N^2 r  \over  g}  \left( \delta \tilde {\mathcal U}_j - \delta \tilde \Phi_j - { g \over r} \tilde W_j \right)\ ,
\end{multline}
so,
for $j=l'+1$ and making use of \eqref{duplus} we have
\begin{equation}
 r\partial_r \delta \tilde{  \mathcal U}_{l'+1}  - \left[ (l'+1) + { \mathcal N^2 r  \over  g} \right]   \delta \tilde {\mathcal U}_{l'+1}
 =  - { \mathcal N^2 r  \over  g}  \left(  \delta \tilde \Phi_{l'+1}+ { g \over r} \tilde W_{l'+1} \right)\ .
\end{equation}
It also follows that 
\begin{equation}
    \delta \tilde \Phi_{l'+3} + { g \over r} \tilde W_{l'+3} = 0 \ .
\end{equation}
Similarly, with $j=l'-1$ we get 
\begin{equation}
 r\partial_r \delta \tilde{  \mathcal U}_{l'-1} + \left[ l' - { \mathcal N^2 r  \over  g} \right]   \delta \tilde {\mathcal U}_{l'-1}
  =  - { \mathcal N^2 r  \over  g}  \left(   \delta \tilde \Phi_{l'-1}+ { g \over r} \tilde W_{l'-1} \right) 
\end{equation}
and we also have 
\begin{equation}
    \delta \tilde \Phi_{l'-3} + { g \over r} \tilde W_{l'-3} = 0\ . 
\end{equation}

Combining the results, we see that we must have
$\tilde W_{l'\pm 3} = \delta \tilde p_{l'\pm 3} = \delta \tilde \Phi_{l'\pm 3} = 0$.
Finally, the continuity equation leads to
$\tilde V_{l'\pm 3} = 0$,
while $\delta \tilde \rho_{l'\pm 3} = 0$
follows from the equation of state relation.
Is essence, all polar $l'\pm3$ multipole contributions will vanish. In turn, this means that a general r-mode must truncate with the $\tilde U_{l'\pm 2}$ terms obtained from \eqref{jlp2} and \eqref{jlm2}. This accords with the discussion in \citet{1981Ap&SS..78..483S} and \citet{1983A&A...125..193S}.

\subsection{The $l'=m$ modes}

Having written down the equations we need to solve to determine the frequency correction $\tilde \omega_2$ and the  multipole structure of an r-mode to order $\Omega^2$---notably without any simplifying assumptions other than neglecting the rotational change in shape of the star---we have a decision to make. Do we want to consider a model that is as ``realistic'' as possible---which will require a numerical solution---or are we more focused on the formal structure of the mode solution? The initial answer is quite simple. As we are not including the rotational shape change it is natural to focus on the qualitative nature of the solution. This leads us to the question of which further simplifying assumptions we may consider. 

As already advertised, we are now going to make the Cowling approximation. That is, we assume that $\delta \tilde \Phi_l=0$.
For the problem at hand, this is pragmatic (as we are focusing on qualitative aspects)  and reasonable (as we do not have to solve the Poisson equation for the perturbed gravitational potential). We want to keep the problem tractable enough that we may proceed to solve it by analytic means. We also know from available numerical results that the r-modes are determined with reasonable precision within this approximation (at least in the context of Newtonian gravity).
In the Cowling approximation, we have
\begin{equation}
    \delta \tilde {\mathcal U}_{l\pm 1} = {\delta \tilde p_{l\pm1} \over \rho} \ .
    \label{dUt}
\end{equation}

In the $l'=m$ case, we have
 $\mathcal Q_{l'=m} = 0$ which means that we only need to consider the coupling between the leading order $U_{l'}$ and the polar $l'+1$
 contributions. (The axial second order contribution $\tilde U_{l'+2}$ can be calculated at a second stage.)
 The set of equations to consider now are: (i) the continuity equation \eqref{continu1}
\begin{equation}
     {1\over r^2}   \partial_r \left(r\rho  \tilde W_{l'+1} \right) - {(l'+1)(l'+2) \rho \over r^2} \tilde V_{l'+1} = - {1\over c_s^2} \delta\tilde  p_{l'+1} + {\mathcal N^2 \over g^2} \left( \delta  \tilde p_{l'+1} -  {\rho g \over r} \tilde W_{l'+1} \right) \ ,
     \label{continu2}
\end{equation}
(ii) the differential equation \eqref{Uder1}
\begin{equation}
     r \partial_r U_{l'} - (l'+1) U_{l'}
    =-  {(l'+1)^2 \over 4m \mathcal Q_{l'+1}} {r\over \rho g} {\mathcal N^2 \over \Omega_0^2} \left( \delta \tilde p_{l'+1} -  {\rho g \over r} \tilde W_{l'+1} \right) \ , 
    \label{rmodeq}
\end{equation}
along with (iii) the algebraic relation \eqref{jl}
\begin{equation}
    {(l'+2) \rho \over r} \tilde V_{l'+1} = {(l'+1)\over 2 \mathcal Q_{l'+1}} {\rho \over r} \tilde \omega_2  U_{l'}  -  {\rho \over r} \tilde W_{l'+1}\ , 
\end{equation}
(iv) the relation \eqref{duplus}
\begin{equation}
  {\delta \tilde p_{l'+1} \over \rho } = -{4m  \Omega_0^2 \over (l'+1)^2} \mathcal Q_{l'+1} U_{l'}
\end{equation}
and the surface boundary condition \eqref{surfcon} (obviously). It is easy to see that we end up with two coupled first-order equations for $U_{l'}$ and $\tilde W_{l'+1}$.  

It is instructive to introduce  $U_{l'} = r^{l'+1} \bar U_{l'}$
and rewrite \eqref{rmodeq} as
\begin{equation}
     r \partial_r \bar U_{l'} 
   =   {r\over \rho g} \mathcal N^2 \bar U_{l'} 
   + {(l'+1)^2 \over 4m \mathcal Q_{l'+1}}  {\mathcal N^2 \over \Omega_0^2}  \tilde W_{l'+1} \ . 
   \label{rbareq}
\end{equation}
It follows immediately that, in the barotropic limit (when $\mathcal N^2 \to 0$ for a fixed $\Omega$), we must have
\begin{equation}
    \bar U_{l'} = \mbox{constant} \Longrightarrow U_{l'}= r^{l'+1} \ .
    \label{rsol}
\end{equation}
The only alternative would be for 
\begin{equation}
   {\mathcal N^2 \over \Omega_0^2} \tilde W_{l'+1} =  {\mathcal N^2 \over \Omega^2} W_{l'+1}  
\end{equation}
to remain finite in the barotropic limit. However, this would violate the assumed slow-rotation ordering for the r-mode solution. That this happens should not be a surprise given the general discussion in Section~\ref{lowfreq}. The result is simply an illustration of the fact that we need to make different assumption in barotropic regions of the star.

In general, we need to solve \eqref{rbareq} along with the continuity equation \eqref{continu2}, which becomes
\begin{equation}
     {1\over r^2}  \partial_r \left(r\rho  \tilde W_{l'+1} \right) + \left[ {(l'+1) \over r^2} + {\mathcal N^2 \over rg}\right] \rho \tilde W_{l'+1}
     = \left[ {(l'+1)^2 \over 2\mathcal Q_{l'+1}^2 }{\tilde \omega_2 \over r^2}
      -{4m  \Omega_0^2 \over (l'+1)^2} {1\over c_s^2} \left( 1 - {\mathcal N^2 c_s^2 \over g^2} \right)   \right] \rho \mathcal Q_{l'+1}  r^{l'+1} \bar U_{l'} \ .
\end{equation}
The two equations (plus the boundary conditions) constitute a Sturm-Liouville problem \citep{1981A&A....94..126P} so we expect to have an infinite set of eigenvalues (see \citet{1982ApJ...256..717S} and \citet{FandA} for indicative results for the r-mode overtones). However, as we have seen, the problem changes in barotropic limit. The overtones disappear and we are left with a single r-mode, represented by \eqref{rsol}.

As a simple example of the single remaining r-mode, we may consider an incompressible barotropic star, for which  $\rho =$ constant so $c_s^2\to \infty$ and we are left with
\begin{equation}
      \partial_r \left(r^{l'+2}  \tilde W_{l'+1} \right)
     =  {(l'+1)^2 \over 2\mathcal Q_{l'+1} }\tilde \omega_2   r^{2(l'+1)} \bar U_{l'} \ ,
\end{equation}
with $\bar U_{l'} = A = $~constant. This leads to
\begin{equation}
    \tilde W_{l'+1}
        =  {(l'+1)^2 \over 2 (2l'+3)  \mathcal Q_{l'+1}} \tilde \omega_2   A r^{l'+1} 
\end{equation}
and we also have
\begin{equation}
   \delta \tilde p_{l'+1} = -{4m  \Omega_0^2 \over (l'+1)^2} \mathcal Q_{l'+1} \rho r^{l'+1} A \ .
\end{equation}
Finally, the surface boundary condition becomes
\begin{equation}
     -{4m   \over (l'+1)^2} \mathcal Q_{l'+1}   -   {(l'+1)^2 \over 2 (2l'+3)  \mathcal Q_{l'+1}} \tilde \omega_2  = 0 \ ,
\end{equation}
so we arrive at
\begin{equation}
    \tilde \omega_2 = - { 8m \over (l'+1)^4}\ , 
    \label{omeg2}
\end{equation}
since 
\begin{equation}
    Q_m^2 = {1\over 2m+3} \ .
\end{equation}
when $l'=m$. We briefly  compare this result to available results from the literature in Appendix~\ref{app}.

In summary, our arguments clearly  illustrate the known fact that the nature of the $l'=m$ r-mode problem changes as stratification weakens. The evidence is clear. We have to approach the $\mathcal N^2 \to 0$ limit with care. In fact, for neutron stars the problem is particularly intricate. Taking the results in Figure~\ref{epshat} at face value, we always have to assume the region close to the surface of the star to be barotropic, while the high-density region may not be. This, in turn, means that the assumed slow-rotation ordering  associated with the non-barotropic r-mode must break (as $\hat \epsilon \lesssim \epsilon$ close to the star's surface). In effect,  the formulation of the problem---as we have presented it---is not consistent. This presents a technical challenge as the solution needs to smoothly join the stratified region where the stratified assumptions hold with a barotropic region where the equation becomes those associated with the general inertial modes. As far as we are aware, this problem has not been considered (at least not for neutron stars), although the required strategy---basically abandoning the slow-rotation ordering for the perturbation---has been developed and employed to good effect for main sequence stars \citep{2006MNRAS.365..677L}. 
 
\subsection{The $l'\neq m$ modes}

Let us now turn to the $l'\neq m$ r-modes. In general, we then need to consider 
the continuity equations for the polar $l'\pm 1$ contributions. In particular, we need to solve the 
differential equations \eqref{Uder1} and \eqref{Uder2}:
\begin{equation}
     r \partial_r U_{l'} - (l'+1) U_{l'}
    =-  {(l'+1)^2 \over 4m \mathcal Q_{l'+1}} {r\over \rho g} {\mathcal N^2 \over \Omega_0^2} \left( \delta \tilde p_{l'+1} -  {\rho g \over r} \tilde W_{l'+1} \right)  
    \label{dupl}
\end{equation}
and
\begin{equation}
     r\partial_r U_{l'} + l' U_{l'}
   = - {(l')^2 \over 4m \mathcal Q_{l'}} {r\over \rho g} {\mathcal N^2 \over \Omega_0^2} \left( \delta \tilde p_{l'-1} -  {\rho g \over r} \tilde W_{l'-1} \right) \ .
   \label{dumi}
\end{equation}  
It is easy to see that we run into trouble in the barotropic limit. Combining \eqref{dupl} and \eqref{dumi} we have an algebraic relation:
\begin{equation}
    (2l'+1) U_{l'} =  - {1\over 4m} {\mathcal N^2 \over \Omega_0^2}   \left[ {(l'+1)^2 \over \mathcal Q_{l'+1}} \tilde W_{l'+1} - {(l')^2 \over \mathcal Q_{l'}} \tilde W_{l'-1} \right] \ ,
\end{equation}
or, alternatively, 
\begin{equation}
   \tilde W_{l'-1} = {\mathcal Q_{l'} \over (l')^2} \left[  (2l'+1) 4m  {\Omega_0^2 \over \mathcal N^2}  U_{l'} +  {(l'+1)^2 \over \mathcal Q_{l'+1}} \tilde W_{l'+1}  \right] \ .
\end{equation}
The first relation shows that, if $\tilde W_{l'\pm 1}$ remain finite then we must have $U_{l'}\to 0$ in the barotropic limit. This would be incompatible with \eqref{dupl}. We get a hint of the resolution to the problem from the alternative version, which suggests that if we insist that $U_{l'}$ remains $\mathcal O(1)$ when $\mathcal N^2\to 0$ then $\tilde W_{l'-1}$ must diverge. 

The unavoidable  conclusion is that the $l'\neq m$ r-modes cannot exist in the barotropic limit---as expected from the arguments by \citet{1999ApJ...521..764L}. If $\mathcal N^2\to 0 $ for a fixed rotation rate $\Omega$, then the assumed r-mode ordering must break. In fact, if $\mathcal N^2=0$ at some point in the star we have a problem. Given the available equations there does not seem to be a way to avoid dividing by $\mathcal N^2$ so the problem will be singular.

In summary, while the stratified problem can be solved for $l'\neq m$ r-modes \citep{1982ApJ...256..717S}, the solution does not apply for realistic neutron star models, see \citet{FandA} for related numerical results. If we want to consider the actual problem, then we have to rethink our strategy. This again suggests that we may need to abandon the slow-rotation ordering a tackle the general problem numerically (as in the body of work by Lee and collaborators \citep{1987MNRAS.224..513L,1995A&A...301..419L,1997ApJ...491..839L,2000ApJS..129..353Y,2000ApJ...529..997Y,2006MNRAS.365..677L}).

\section{A physically motivated alternative}
\label{physmot}

Based on the stratification results from Figure~\ref{epshat}, it makes sense---for the fastest spinning stars---to consider the stratification to be second order in the slow rotation expansion. We then have \eqref{neweps} which leads to (at order $\epsilon^2$)
\begin{equation}
c_s^2 \delta \tilde \rho_j  =   \delta \tilde p_j + {\mathcal N^2 c_s^2 \rho  \over r g} W_j  \ ,
\end{equation}
where $W_j\sim \mathcal O(1)$. This suggests that we change the assumed ordering in such a way that  $W_j \to W_j + \epsilon^2 \tilde W_j$
and similar for $V_j$. The axial displacement remains as before. With the polar displacement components present already at leading order, the problem is close to that for a general inertial mode. 

With these assumptions, the leading order continuity equation requires
\begin{equation}
     {1 \over r^2} \partial_r \left( r\rho W_j\right)   - j(j+1) {\rho \over r^2}  V_j = 0\ .
\end{equation}

The radial vorticity equation leads to, at order $\epsilon$:
\begin{equation}
 \left[ j(j+1) \omega_0  - 2m \right] U_j + 2 (j+1) [ W_{j-1} - (j-1) V_{j-1}]  \mathcal Q_{j}
 - 2 j[ W_{j+1} + (j+2) V_{j+1}] \mathcal Q_{j+1} = 0   \ .
\end{equation}

It is also convenient to use the algebraic relation from the divergence equation, which at order $\epsilon^2$ provides :
\begin{equation}
 \omega_0 \left[ j(j+1) \omega_0  -2m\right] V_j -2m \omega_0  W_j  - j(j+1)  \delta  \tilde {\mathcal U}_j   
 -  2\omega_0 \left[  (j-1)(j+1) \mathcal Q_{j} U_{j-1} +  j (j+2) \mathcal Q_{j+1}  U_{j+1} \right] = 0 \ .
\end{equation}

Finally, in this case it seems natural (given that the horizontal vorticity equation involves derivatives of all three displacement components 
to use the radial Euler equation, which leads to, at order $\epsilon^2$:
\begin{multline}
 - \omega_0^2 W_j +   2  m \omega_0  V_j
+ r\partial_r \delta  \tilde{\mathcal U_j}  
+ 2 \omega_0 \left[  (j-1) \mathcal Q_j U_{j-1}  - (j+2) \mathcal Q_{j+1} U_{j+1} \right] \\
= - { r \over \rho^2 \Omega_0^2} \left( \delta \tilde p_j \partial_r \rho - \delta  \tilde \rho_j  \partial_r  p\right) =  - {r \over \rho H \Omega_0^2} ( c_s^2 \delta \tilde \rho_j - \delta \tilde p_j) = - {\mathcal N^2 c_s^2   \over  g H \Omega_0^2} W_j = {\mathcal N^2 \over \Omega_0^2}   W_j \ .
\end{multline}
Finally, in the Cowling approximation we have \eqref{dUt} and we also need \eqref{angvor}, which at order $\epsilon^2$ leads to 
\begin{multline}
    \omega_0 \left[2  \left(1 - \mathcal Q_j^2 - \mathcal Q_{j+1}^2 \right) r\partial_r W_j + m\omega_0 W_j \right] 
    \\
     -\omega_0 \left[  \left\{ m\omega_0 +2\left[ (j+1) \mathcal Q_j^2- j\mathcal Q_{j+1}^2 \right]  \right\}  r\partial_r V_j + 2m^2 V_j \right] \\
    - \omega_0 \left[ \left( \omega_0 (j-1)  - 2m \right) r\partial_r U_{j-1}  +2m (j-1)   U_{j-1} \right] \mathcal Q_{j}   \\
+\omega_0  \left\{ \left[ \omega_0 (j+2)  + 2m \right] r\partial_r U_{j+1}  + 2m (j+2)  U_{j+1} \right\} \mathcal Q_{j+1}\\
+ 2\omega_0  \left[ (j-2) r\partial_r V_{j-2} - r \partial_r W_{j-2} \right] \mathcal Q_{j-1} \mathcal Q_{j}  - 2\omega_0 \left[ (j+3) r\partial_r V_{j+2} + r \partial_r W_{j+2} \right] \mathcal Q_{j+2} \mathcal Q_{j+1}  \\ = -  {mr\over \rho^2 \Omega_0^2} \left(\delta \tilde \rho_j \partial_r p - \delta  \tilde p_j \partial_r \rho \right) = - {m  \mathcal N^2 \over \Omega_0^2} W_j \ .
\label{KLcompare}
\end{multline}

Let us focus on the problem we would have to solve in order  to identify a solution ``close to'' the traditional r-mode. That is, we are looking for modes such that $U_{l'}\sim\mathcal O(1)$ with $l'=m$ and with a frequency given by \eqref{omex} and \eqref{rfreq}. With the usual ordering for stratified stars, this would include the r-mode overtones. The only difference here is that we are no longer (necessarily) assuming that the polar displacement contributions enter at higher slow-rotation order. The equations that involve $U_{l'}$ are then, first of all, the leading order relation
\begin{equation}
 \left[ l'(l'+1) \omega_0  - 2m \right] U_{l'}
 - 2 l'[ W_{l'+1} + (l'+2) V_{l'+1}] \mathcal Q_{l'+1}= 0  \ ,
\end{equation}
which, for a mode with the usual leading order r-mode frequency \eqref{rfreq}, reduces to 
\begin{equation}
 - 2 l'[ W_{l'+1} + (l'+2) V_{l'+1}] \mathcal Q_{l'+1}=  0 \ . 
\end{equation}
This can be combined with the leading order continuity equation to give 
\begin{equation}
     \partial_r \left( \rho W_{l'+1}\right)   + (l'+2) \rho   W_{l'+1} = 0\ , 
\end{equation}
which leads to, with $A$ constant,
\begin{equation}
r^{l'+2}\rho W_{l'+1} = A \Longrightarrow W_{l'+1} = \frac{A}{\rho r^{l'+2}}\ .
\end{equation}
This is clearly problematic as the solution diverges at the centre of the star (and at the surface as well, if $\rho \to 0$ as $r\to R$). The only way to avoid trouble is to have the trivial solution, $A=0$, and move on to the equations for the higher order terms. If we do this then we immediately see that the problem is \emph{identical} to the one we already solved for barotropic stars. There will be a single r-mode for each $l'=m$. This tells us that the frequencies of the r-mode overtones can no longer be given by \eqref{omex}. As expected---and in accordance with the results from Figure~4 of \citet{2000ApJS..129..353Y}---they have to change character.

Next, for $l'\neq m$ it is easy to see that the equations we are now considering still lead to a singular problem \emph{unless} the polar components $W_j$ and $V_j$ are $\mathcal O(1)$.
This is as expected: We need to consider solutions close to the barotropic inertial modes.

Finally, for the  modes of the fastest spinning neutron stars we see that the problem (to leading order) is very close to the inertial-mode problem as formulated by \citet{1999ApJ...521..764L}. The only difference is the right-hand side of \eqref{KLcompare}. In essence, the problem we need to consider if we want to establish the astrophysical role of the gravitational-wave driven r-mode instability is close, but not identical, to the inertial-mode problem. As far as we are aware, this problem has not been stated despite the numerous discussions of the r-mode instability in the literature. This problem clearly needs further attention and our intention is to approach it numerically (also accounting for the rotational shape corrections, following the strategy outlined in \citet{FandA}) in the near future.

\section{Conclusions and outlook}

We have revisited the problem of inertial r-modes in stratified neutron stars. Our motivation for this was two-fold. 
First, we wanted to add realism to the discussion by introducing a more precise description of the composition stratification in a mature neutron star. Our analysis of the problem highlights issues with the traditional approach to the problem. In order to account for the expected variation of the internal composition stratification with density, we need to rethink the computational strategy for determining the r-modes. There appears to be two strategies for dealing with this problem. The first would simply involve introducing the standard slow-rotation expansion for the perturbation and approach the problem numerically from the outset. This approach has been championed in a series of papers (albeit not for realistic neutron star stratification) by Umin Lee and colleagues \citep{1987MNRAS.224..513L,1995A&A...301..419L,1997ApJ...491..839L,2000ApJS..129..353Y,2000ApJ...529..997Y,2006MNRAS.365..677L}. Our discussion suggests this may be the only viable alternative for moderate to slowly rotating neutron stars. An alternative approach would be to focus on the fastest (known) spinning stars. For these, the stratification is expected to be relatively weak and the slow-rotation expansion is (again) viable. We have shown that this leads to a problem close to that for inertial modes, as formulated by \citet{1999ApJ...521..764L}.

This brings us to our second---somewhat deeper---motivation. We wanted to shed light on the (still unresolved) problem of r-modes in stratified relativistic stars \citep{1998MNRAS.293...49K,1999ApJ...520..788K,1999MNRAS.308..745B,2004CQGra..21.4661L,2002ApJ...567.1112Y,2005MNRAS.363..121P,2021arXiv211201171K,2022Univ....8..542K}. In this context, our analysis also suggests issues with the standard formulation of the problem. We expect that the long-standing issue of a singularity associated with internal co-rotation points will be resolved once the r-mode problem is reformulated as a generalised inertial mode problem (in the spirit of the discussion in Section~\ref{physmot} above). This is likely to lead to mode-solutions fairly close to the inertial modes and hence results similar to those of \citet{2000PhRvD..63b4019L,2003PhRvD..68l4010L,2003MNRAS.339.1170R} and \citet{2015PhRvD..91b4001I}. The latter may be particularly important as the modes are determined for realistic (barotropic) equations of state. If it turns out to be the case that the mode solutions shift only slightly once we account for the stratification then the result we need for (say) gravitational-wave searches may already be at hand. Of course, at this point this is speculation. There are calculations to be done in order to verify the assertion.
In addition, the implications of our discussion for the range of problems where the r-modes are thought to play a role, from limiting the spin of neutron stars to the dynamical tide in a neutron-star binary, also remain to be explored.

\begin{acknowledgements}
We acknowledge support from STFC via grant number ST/V000551/1. The contribution of NA was partly carried out at the Aspen Center for Physics, which is supported by National Science Foundation grant PHY-1607611. He also thanks the Simons Foundation for generous travel support. The contribution of FG was partly carried out at the Institute for Nuclear Theory at the University of Washington during the ``Neutron Rich Matter on Heaven and Earth'' workshop, which is supported by the U.S. Department of Energy grant DE-FG02-00ER41132.
\end{acknowledgements}

\appendix
\section{Comparing to the literature}
\label{app}

As a slight addendum to our discussion, we worked out the order $\Omega^2$ frequency correction for the single r-mode that remains (for $l'=m$) in barotropic stars. It is interesting to compare the result to similar results in the literature. From  equation~\eqref{omeg2} we have (with  $l'=m$)
\begin{equation}
    \tilde \omega_2 = - { 8m \over (m+1)^4}\ .
    \label{om2f}
\end{equation}
As our calculation assumed constant density, it is natural to first compare to the results from \cite{1999A&A...341..110K}. Working with the equations from \citet{1982ApJ...256..717S}, hence including the rotational change in shape, they  arrive at (with our conventions)
\begin{equation}
    \tilde \omega_2  = {5m\over (m+1)^2} \ .
    \label{ksresult}
\end{equation}
Evidently, the result is different from ours. The first clue to the origin of the difference comes once we note that the \cite{1999A&A...341..110K} frequency correction vanishes if we ignore the rotational change in shape. This suggests that we are not comparing like for like. This becomes apparent when we turn to the results from the Appendix of \citet{1981A&A....94..126P}. They have 
\begin{equation}
    \tilde \omega_2 = - {8 \over (m+1)^4} + {5m\over (m+1)^2}\ .
\end{equation}
Here we recognise the second term as the result from \cite{1999A&A...341..110K}. In essence, this is the rotationally induced frequency correction.
This seems quite intuitive. Of course, the first term from the \citet{1981A&A....94..126P} result still differs from ours (there is a missing factor of $m$). While we have not been able to pinpoint the origin of the discrepancy, we have reworked the calculation from \citet{1981A&A....94..126P} and the result we get accords with \eqref{om2f}.

For constant density stars it turns out to be straightforward to account for the rotational change in shape. Following the strategy from \citet{1982ApJ...256..717S} we note that that shape correction only impacts on the surface boundary condition. Working this out, we arrive at (still in the Cowling approximation)
\begin{equation}
   \tilde  \omega_2  = - {8m \over (m+1)^4} + {5m \over (m+1)^2} \ .
   \label{provres}
\end{equation}
This agree with the identification of \eqref{ksresult} as the rotational correction to the r-mode. 

It is also relativity easy to relax the Cowling approximation. Again, for constant density stars this only affects the surface boundary condition. 
Working this out, we find that the inclusion of the perturbed gravitational potential adds a multiplicative factor to the frequency correction. Instead of \eqref{om2f} we get
\begin{equation}
    \tilde \omega_2 = - {8m \over (m+1)^4}  \left( {2m+3 \over 2m} \right)\ .
\end{equation}
Now,  \citet{1981A&A....94..126P} state that their result follows if one expands the result from \citet{1889RSPTA.180..187B} in spherical harmonics. This assertion is difficult to confirm, but it is supported by the result from \citet{1999PhRvD..59d4009L}. Their result leads to 
\begin{equation}
    \tilde \omega_2  = - { 4(2m+3) \over m (m+1)^4} + {5m\over (m+1)^2}\ .
\end{equation}
Again,  we recognize the shape correction. Assuming that the factor due to the Cowling approximation is the one we determined, the result agrees with \eqref{provres}. This still leaves us with the missing factor of $m$ compared to \eqref{om2f}. 

In essence, the results in the available literature are not quite consistent. Having said that, as we have checked our calculation leading to \eqref{om2f} several times, we stand by this as the correct result.

\bibliography{bibliography}

\begin{thebibliography}{}
\expandafter\ifx\csname natexlab\endcsname\relax\def\natexlab#1{#1}\fi
\providecommand{\url}[1]{\href{#1}{#1}}
\providecommand{\dodoi}[1]{doi:~\href{http://doi.org/#1}{\nolinkurl{#1}}}
\providecommand{\doeprint}[1]{\href{http://ascl.net/#1}{\nolinkurl{http://ascl.net/#1}}}
\providecommand{\doarXiv}[1]{\href{https://arxiv.org/abs/#1}{\nolinkurl{https://arxiv.org/abs/#1}}}

\bibitem[{{Aasi} {et~al.}(2015)}]{2015ApJ...813...39A}
{Aasi}, J., {et~al.} 2015, Ap. J., 813, 39, \dodoi{10.1088/0004-637X/813/1/39}

\bibitem[{{Abadie} {et~al.}(2010)}]{2010ApJ...722.1504A}
{Abadie}, J., {et~al.} 2010, Ap. J., 722, 1504,
  \dodoi{10.1088/0004-637X/722/2/1504}

\bibitem[{{Abbott} {et~al.}(2021{\natexlab{a}})}]{2021ApJ...921...80A}
{Abbott}, R., {et~al.} 2021{\natexlab{a}}, Ap. J., 921, 80,
  \dodoi{10.3847/1538-4357/ac17ea}

\bibitem[{{Abbott} {et~al.}(2021{\natexlab{b}})}]{2021ApJ...922...71A}
---. 2021{\natexlab{b}}, Ap. J., 922, 71, \dodoi{10.3847/1538-4357/ac0d52}

\bibitem[{{Abbott} {et~al.}(2022{\natexlab{a}})}]{2022PhRvD.105b2002A}
---. 2022{\natexlab{a}}, Phys. Rev. D, 105, 022002,
  \dodoi{10.1103/PhysRevD.105.022002}

\bibitem[{{Abbott} {et~al.}(2022{\natexlab{b}})}]{2022PhRvD.106d2003A}
---. 2022{\natexlab{b}}, Phys. Rev. D, 106, 042003,
  \dodoi{10.1103/PhysRevD.106.042003}

\bibitem[{{Alford} \& {Harris}(2018)}]{2018PhRvC..98f5806A}
{Alford}, M.~G., \& {Harris}, S.~P. 2018, Phys. Rev. C, 98, 065806,
  \dodoi{10.1103/PhysRevC.98.065806}

\bibitem[{{Andersson}(1998)}]{1998ApJ...502..708A}
{Andersson}, N. 1998, Ap. J., 502, 708, \dodoi{10.1086/305919}

\bibitem[{{Andersson}(2003)}]{2003CQGra..20R.105A}
---. 2003, Classical and Quantum Gravity, 20, R105,
  \dodoi{10.1088/0264-9381/20/7/201}

\bibitem[{{Andersson} {et~al.}(2014){Andersson}, {Jones}, \&
  {Ho}}]{2014MNRAS.442.1786A}
{Andersson}, N., {Jones}, D.~I., \& {Ho}, W.~C.~G. 2014, MNRAS, 442, 1786,
  \dodoi{10.1093/mnras/stu870}

\bibitem[{{Andersson} {et~al.}(1999{\natexlab{a}}){Andersson}, {Kokkotas}, \&
  {Schutz}}]{1999ApJ...510..846A}
{Andersson}, N., {Kokkotas}, K., \& {Schutz}, B.~F. 1999{\natexlab{a}}, Ap. J.,
  510, 846, \dodoi{10.1086/306625}

\bibitem[{{Andersson} \& {Kokkotas}(2001)}]{2001IJMPD..10..381A}
{Andersson}, N., \& {Kokkotas}, K.~D. 2001, International Journal of Modern
  Physics D, 10, 381, \dodoi{10.1142/S0218271801001062}

\bibitem[{{Andersson} {et~al.}(1999{\natexlab{b}}){Andersson}, {Kokkotas}, \&
  {Stergioulas}}]{1999ApJ...516..307A}
{Andersson}, N., {Kokkotas}, K.~D., \& {Stergioulas}, N. 1999{\natexlab{b}},
  Ap. J., 516, 307, \dodoi{10.1086/307082}

\bibitem[{{Andersson} \& {Pnigouras}(2019)}]{2019MNRAS.489.4043A}
{Andersson}, N., \& {Pnigouras}, P. 2019, MNRAS, 489, 4043,
  \dodoi{10.1093/mnras/stz2449}

\bibitem[{{Arras} {et~al.}(2003){Arras}, {Flanagan}, {Morsink}, {Schenk},
  {Teukolsky}, \& {Wasserman}}]{2003ApJ...591.1129A}
{Arras}, P., {Flanagan}, E.~E., {Morsink}, S.~M., {et~al.} 2003, Ap. J., 591,
  1129, \dodoi{10.1086/374657}

\bibitem[{{Beyer} \& {Kokkotas}(1999)}]{1999MNRAS.308..745B}
{Beyer}, H.~R., \& {Kokkotas}, K.~D. 1999, MNRAS, 308, 745,
  \dodoi{10.1046/j.1365-8711.1999.02739.x}

\bibitem[{{Bondarescu} {et~al.}(2007){Bondarescu}, {Teukolsky}, \&
  {Wasserman}}]{2007PhRvD..76f4019B}
{Bondarescu}, R., {Teukolsky}, S.~A., \& {Wasserman}, I. 2007, Phys. Rev. D,
  76, 064019, \dodoi{10.1103/PhysRevD.76.064019}

\bibitem[{{Bondarescu} {et~al.}(2009){Bondarescu}, {Teukolsky}, \&
  {Wasserman}}]{2009PhRvD..79j4003B}
---. 2009, Phys. Rev. D, 79, 104003, \dodoi{10.1103/PhysRevD.79.104003}

\bibitem[{{Bondarescu} \& {Wasserman}(2013)}]{2013ApJ...778....9B}
{Bondarescu}, R., \& {Wasserman}, I. 2013, Ap. J., 778, 9,
  \dodoi{10.1088/0004-637X/778/1/9}

\bibitem[{{Brink} {et~al.}(2004{\natexlab{a}}){Brink}, {Teukolsky}, \&
  {Wasserman}}]{2004PhRvD..70l4017B}
{Brink}, J., {Teukolsky}, S.~A., \& {Wasserman}, I. 2004{\natexlab{a}}, Phys.
  Rev. D, 70, 124017, \dodoi{10.1103/PhysRevD.70.124017}

\bibitem[{{Brink} {et~al.}(2004{\natexlab{b}}){Brink}, {Teukolsky}, \&
  {Wasserman}}]{2004PhRvD..70l1501B}
---. 2004{\natexlab{b}}, Phys. Rev. D, 70, 121501,
  \dodoi{10.1103/PhysRevD.70.121501}

\bibitem[{{Brink} {et~al.}(2005){Brink}, {Teukolsky}, \&
  {Wasserman}}]{2005PhRvD..71f4029B}
---. 2005, Phys. Rev. D, 71, 064029, \dodoi{10.1103/PhysRevD.71.064029}

\bibitem[{{Bryan}(1889)}]{1889RSPTA.180..187B}
{Bryan}, G.~H. 1889, Philosophical Transactions of the Royal Society of London
  Series A, 180, 187, \dodoi{10.1098/rsta.1889.0006}

\bibitem[{{Covas} {et~al.}(2022){Covas}, {Papa}, {Prix}, \&
  {Owen}}]{2022ApJ...929L..19C}
{Covas}, P.~B., {Papa}, M.~A., {Prix}, R., \& {Owen}, B.~J. 2022, Ap. J. Lett.,
  929, L19, \dodoi{10.3847/2041-8213/ac62d7}

\bibitem[{{Fantina} {et~al.}(2013){Fantina}, {Chamel}, {Pearson}, \&
  {Goriely}}]{2013A&A...559A.128F}
{Fantina}, A.~F., {Chamel}, N., {Pearson}, J.~M., \& {Goriely}, S. 2013, \aap,
  559, A128, \dodoi{10.1051/0004-6361/201321884}

\bibitem[{{Fesik} \& {Papa}(2020)}]{2020ApJ...895...11F}
{Fesik}, L., \& {Papa}, M.~A. 2020, Ap. J., 895, 11,
  \dodoi{10.3847/1538-4357/ab8193}

\bibitem[{{Flanagan} \& {Racine}(2007)}]{2007PhRvD..75d4001F}
{Flanagan}, {\'E}.~{\'E}., \& {Racine}, {\'E}. 2007, Phys. Rev. D, 75, 044001,
  \dodoi{10.1103/PhysRevD.75.044001}

\bibitem[{{Friedman} \& {Morsink}(1998)}]{1998ApJ...502..714F}
{Friedman}, J.~L., \& {Morsink}, S.~M. 1998, Ap. J, 502, 714,
  \dodoi{10.1086/305920}

\bibitem[{Gaertig {et~al.}(2011)Gaertig, Glampedakis, Kokkotas, \&
  Zink}]{PhysRevLett.107.101102}
Gaertig, E., Glampedakis, K., Kokkotas, K.~D., \& Zink, B. 2011, Phys. Rev.
  Lett., 107, 101102, \dodoi{10.1103/PhysRevLett.107.101102}

\bibitem[{Gaertig \& Kokkotas(2009)}]{PhysRevD.80.064026}
Gaertig, E., \& Kokkotas, K.~D. 2009, Phys. Rev. D, 80, 064026,
  \dodoi{10.1103/PhysRevD.80.064026}

\bibitem[{Gaertig \& Kokkotas(2011)}]{PhysRevD.83.064031}
---. 2011, Phys. Rev. D, 83, 064031, \dodoi{10.1103/PhysRevD.83.064031}

\bibitem[{{Gittins} \& {Andersson}(2022)}]{FandA}
{Gittins}, F., \& {Andersson}, N. 2022.
\newblock \doarXiv{2212.04892}

\bibitem[{{Glampedakis} \& {Andersson}(2006)}]{2006PhRvD..74d4040G}
{Glampedakis}, K., \& {Andersson}, N. 2006, Phys. Rev. D, 74, 044040,
  \dodoi{10.1103/PhysRevD.74.044040}

\bibitem[{{Gusakov} \& {Kantor}(2013)}]{2013PhRvD..88j1302G}
{Gusakov}, M.~E., \& {Kantor}, E.~M. 2013, Phys. Rev. D, 88, 101302,
  \dodoi{10.1103/PhysRevD.88.101302}

\bibitem[{{Haensel} {et~al.}(2002){Haensel}, {Levenfish}, \&
  {Yakovlev}}]{2002A&A...394..213H}
{Haensel}, P., {Levenfish}, K.~P., \& {Yakovlev}, D.~G. 2002, \aap, 394, 213,
  \dodoi{10.1051/0004-6361:20021112}

\bibitem[{{Haskell} \& {Schwenzer}(2021)}]{2021arXiv210403137H}
{Haskell}, B., \& {Schwenzer}, K. 2021, arXiv e-prints, arXiv:2104.03137.
\newblock \doarXiv{2104.03137}

\bibitem[{{Hessels} {et~al.}(2006){Hessels}, {Ransom}, {Stairs}, {Freire},
  {Kaspi}, \& {Camilo}}]{2006Sci...311.1901H}
{Hessels}, J.~W.~T., {Ransom}, S.~M., {Stairs}, I.~H., {et~al.} 2006, Science,
  311, 1901, \dodoi{10.1126/science.1123430}

\bibitem[{{Idrisy} {et~al.}(2015){Idrisy}, {Owen}, \&
  {Jones}}]{2015PhRvD..91b4001I}
{Idrisy}, A., {Owen}, B.~J., \& {Jones}, D.~I. 2015, Phys. Rev. D, 91, 024001,
  \dodoi{10.1103/PhysRevD.91.024001}

\bibitem[{Jones {et~al.}(2002)Jones, Andersson, \&
  Stergioulas}]{10.1046/j.1365-8711.2002.05566.x}
Jones, D.~I., Andersson, N., \& Stergioulas, N. 2002, Monthly Notices of the
  Royal Astronomical Society, 334, 933,
  \dodoi{10.1046/j.1365-8711.2002.05566.x}

\bibitem[{{Kojima}(1998)}]{1998MNRAS.293...49K}
{Kojima}, Y. 1998, MNRAS, 293, 49, \dodoi{10.1046/j.1365-8711.1998.01119.x}

\bibitem[{{Kojima} \& {Hosonuma}(1999)}]{1999ApJ...520..788K}
{Kojima}, Y., \& {Hosonuma}, M. 1999, Ap. J., 520, 788, \dodoi{10.1086/307481}

\bibitem[{{Kokkotas} \& {Stergioulas}(1999)}]{1999A&A...341..110K}
{Kokkotas}, K.~D., \& {Stergioulas}, N. 1999, \aap, 341, 110.
\newblock \doarXiv{astro-ph/9805297}

\bibitem[{{Kraav} {et~al.}(2021){Kraav}, {Gusakov}, \&
  {Kantor}}]{2021arXiv211201171K}
{Kraav}, K.~Y., {Gusakov}, M.~E., \& {Kantor}, E.~M. 2021, arXiv e-prints,
  arXiv:2112.01171.
\newblock \doarXiv{2112.01171}

\bibitem[{{Kraav} {et~al.}(2022){Kraav}, {Gusakov}, \&
  {Kantor}}]{2022Univ....8..542K}
---. 2022, Universe, 8, 542, \dodoi{10.3390/universe8100542}

\bibitem[{{Kr{\"u}ger} {et~al.}(2021){Kr{\"u}ger}, {Kokkotas}, {Manoharan}, \&
  {V{\"o}lkel}}]{2021FrASS...8..166K}
{Kr{\"u}ger}, C.~J., {Kokkotas}, K.~D., {Manoharan}, P., \& {V{\"o}lkel}, S.~H.
  2021, Frontiers in Astronomy and Space Sciences, 8, 166,
  \dodoi{10.3389/fspas.2021.736918}

\bibitem[{{Lasky}(2015)}]{2015PASA...32...34L}
{Lasky}, P.~D. 2015, Publ. Astron. Soc. Australia, 32, e034,
  \dodoi{10.1017/pasa.2015.35}

\bibitem[{{Lee}(2006)}]{2006MNRAS.365..677L}
{Lee}, U. 2006, MNRAS, 365, 677, \dodoi{10.1111/j.1365-2966.2005.09751.x}

\bibitem[{{Lee}(2014)}]{2014MNRAS.442.3037L}
---. 2014, MNRAS, 442, 3037, \dodoi{10.1093/mnras/stu1077}

\bibitem[{{Lee} \& {Baraffe}(1995)}]{1995A&A...301..419L}
{Lee}, U., \& {Baraffe}, I. 1995, \aap, 301, 419

\bibitem[{{Lee} \& {Saio}(1987)}]{1987MNRAS.224..513L}
{Lee}, U., \& {Saio}, H. 1987, MNRAS, 224, 513, \dodoi{10.1093/mnras/224.3.513}

\bibitem[{{Lee} \& {Saio}(1997)}]{1997ApJ...491..839L}
---. 1997, Ap. J., 491, 839, \dodoi{10.1086/304980}

\bibitem[{{Lee} \& {Strohmayer}(1996)}]{1996A&A...311..155L}
{Lee}, U., \& {Strohmayer}, T.~E. 1996, \aap, 311, 155

\bibitem[{{Lindblom} \& {Ipser}(1999)}]{1999PhRvD..59d4009L}
{Lindblom}, L., \& {Ipser}, J.~R. 1999, Phys. Rev. D, 59, 044009,
  \dodoi{10.1103/PhysRevD.59.044009}

\bibitem[{{Lindblom} {et~al.}(1998){Lindblom}, {Owen}, \&
  {Morsink}}]{1998PhRvL..80.4843L}
{Lindblom}, L., {Owen}, B.~J., \& {Morsink}, S.~M. 1998, Phys. Rev. Lett., 80,
  4843, \dodoi{10.1103/PhysRevLett.80.4843}

\bibitem[{{Lockitch} {et~al.}(2000){Lockitch}, {Andersson}, \&
  {Friedman}}]{2000PhRvD..63b4019L}
{Lockitch}, K.~H., {Andersson}, N., \& {Friedman}, J.~L. 2000, Phys. Rev. D,
  63, 024019, \dodoi{10.1103/PhysRevD.63.024019}

\bibitem[{{Lockitch} {et~al.}(2004){Lockitch}, {Andersson}, \&
  {Watts}}]{2004CQGra..21.4661L}
{Lockitch}, K.~H., {Andersson}, N., \& {Watts}, A.~L. 2004, Classical and
  Quantum Gravity, 21, 4661, \dodoi{10.1088/0264-9381/21/19/012}

\bibitem[{{Lockitch} \& {Friedman}(1999)}]{1999ApJ...521..764L}
{Lockitch}, K.~H., \& {Friedman}, J.~L. 1999, Ap. J., 521, 764,
  \dodoi{10.1086/307580}

\bibitem[{{Lockitch} {et~al.}(2003){Lockitch}, {Friedman}, \&
  {Andersson}}]{2003PhRvD..68l4010L}
{Lockitch}, K.~H., {Friedman}, J.~L., \& {Andersson}, N. 2003, Phys. Rev. D,
  68, 124010, \dodoi{10.1103/PhysRevD.68.124010}

\bibitem[{{Longuet-Higgins}(1968)}]{1968RSPTA.262..511L}
{Longuet-Higgins}, M.~S. 1968, Philosophical Transactions of the Royal Society
  of London Series A, 262, 511, \dodoi{10.1098/rsta.1968.0003}

\bibitem[{{Ma} {et~al.}(2021){Ma}, {Yu}, \& {Chen}}]{2021PhRvD.103f3020M}
{Ma}, S., {Yu}, H., \& {Chen}, Y. 2021, Phys. Rev. D, 103, 063020,
  \dodoi{10.1103/PhysRevD.103.063020}

\bibitem[{{Owen} {et~al.}(1998){Owen}, {Lindblom}, {Cutler}, {Schutz},
  {Vecchio}, \& {Andersson}}]{1998PhRvD..58h4020O}
{Owen}, B.~J., {Lindblom}, L., {Cutler}, C., {et~al.} 1998, Phys. Rev. D, 58,
  084020, \dodoi{10.1103/PhysRevD.58.084020}

\bibitem[{{Papaloizou} \& {Pringle}(1978)}]{1978MNRAS.182..423P}
{Papaloizou}, J., \& {Pringle}, J.~E. 1978, MNRAS, 182, 423,
  \dodoi{10.1093/mnras/182.3.423}

\bibitem[{{Passamonti} {et~al.}(2016){Passamonti}, {Andersson}, \&
  {Ho}}]{2016MNRAS.455.1489P}
{Passamonti}, A., {Andersson}, N., \& {Ho}, W.~C.~G. 2016, MNRAS, 455, 1489,
  \dodoi{10.1093/mnras/stv2149}

\bibitem[{Passamonti {et~al.}(2009)Passamonti, Haskell, Andersson, Jones, \&
  Hawke}]{10.1111/j.1365-2966.2009.14408.x}
Passamonti, A., Haskell, B., Andersson, N., Jones, D.~I., \& Hawke, I. 2009,
  Monthly Notices of the Royal Astronomical Society, 394, 730,
  \dodoi{10.1111/j.1365-2966.2009.14408.x}

\bibitem[{{Poisson}(2020)}]{2020PhRvD.101j4028P}
{Poisson}, E. 2020, Phys. Rev. D, 101, 104028,
  \dodoi{10.1103/PhysRevD.101.104028}

\bibitem[{{Poisson} \& {Buisson}(2020)}]{2020PhRvD.102j4005P}
{Poisson}, E., \& {Buisson}, C. 2020, Phys. Rev. D, 102, 104005,
  \dodoi{10.1103/PhysRevD.102.104005}

\bibitem[{{Pons} {et~al.}(2005){Pons}, {Gualtieri}, {Miralles}, \&
  {Ferrari}}]{2005MNRAS.363..121P}
{Pons}, J.~A., {Gualtieri}, L., {Miralles}, J.~A., \& {Ferrari}, V. 2005,
  MNRAS, 363, 121, \dodoi{10.1111/j.1365-2966.2005.09429.x}

\bibitem[{{Potekhin} {et~al.}(2013){Potekhin}, {Fantina}, {Chamel}, {Pearson},
  \& {Goriely}}]{2013A&A...560A..48P}
{Potekhin}, A.~Y., {Fantina}, A.~F., {Chamel}, N., {Pearson}, J.~M., \&
  {Goriely}, S. 2013, \aap, 560, A48, \dodoi{10.1051/0004-6361/201321697}

\bibitem[{{Provost} {et~al.}(1981){Provost}, {Berthomieu}, \&
  {Rocca}}]{1981A&A....94..126P}
{Provost}, J., {Berthomieu}, G., \& {Rocca}, A. 1981, \aap, 94, 126

\bibitem[{{Reisenegger} \& {Goldreich}(1992)}]{1992ApJ...395..240R}
{Reisenegger}, A., \& {Goldreich}, P. 1992, Ap. J., 395, 240,
  \dodoi{10.1086/171645}

\bibitem[{{Ruoff} {et~al.}(2003){Ruoff}, {Stavridis}, \&
  {Kokkotas}}]{2003MNRAS.339.1170R}
{Ruoff}, J., {Stavridis}, A., \& {Kokkotas}, K.~D. 2003, MNRAS, 339, 1170,
  \dodoi{10.1046/j.1365-8711.2003.06267.x}

\bibitem[{{Saio}(1981)}]{1981ApJ...244..299S}
{Saio}, H. 1981, Ap. J., 244, 299, \dodoi{10.1086/158708}

\bibitem[{{Saio}(1982)}]{1982ApJ...256..717S}
---. 1982, Ap. J., 256, 717, \dodoi{10.1086/159945}

\bibitem[{{Schenk} {et~al.}(2001){Schenk}, {Arras}, {Flanagan}, {Teukolsky}, \&
  {Wasserman}}]{2001PhRvD..65b4001S}
{Schenk}, A.~K., {Arras}, P., {Flanagan}, {\'E}.~{\'E}., {Teukolsky}, S.~A., \&
  {Wasserman}, I. 2001, Phys. Rev. D, 65, 024001,
  \dodoi{10.1103/PhysRevD.65.024001}

\bibitem[{{Schmitt} \& {Shternin}(2018)}]{2018ASSL..457..455S}
{Schmitt}, A., \& {Shternin}, P. 2018, in Astrophysics and Space Science
  Library, Vol. 457, Astrophysics and Space Science Library, ed. L.~{Rezzolla},
  P.~{Pizzochero}, D.~I. {Jones}, N.~{Rea}, \& I.~{Vida{\~n}a}, 455,
  \dodoi{10.1007/978-3-319-97616-7_9}

\bibitem[{{Smeyers} {et~al.}(1981){Smeyers}, {Craeynest}, \&
  {Martens}}]{1981Ap&SS..78..483S}
{Smeyers}, P., {Craeynest}, D., \& {Martens}, L. 1981, \apss, 78, 483,
  \dodoi{10.1007/BF00648954}

\bibitem[{{Smeyers} \& {Martens}(1983)}]{1983A&A...125..193S}
{Smeyers}, P., \& {Martens}, L. 1983, \aap, 125, 193

\bibitem[{{Strohmayer} \&
  {Mahmoodifar}(2014{\natexlab{a}})}]{2014ApJ...784...72S}
{Strohmayer}, T., \& {Mahmoodifar}, S. 2014{\natexlab{a}}, Ap. J., 784, 72,
  \dodoi{10.1088/0004-637X/784/1/72}

\bibitem[{{Strohmayer} \&
  {Mahmoodifar}(2014{\natexlab{b}})}]{2014ApJ...793L..38S}
---. 2014{\natexlab{b}}, Ap. J. Lett., 793, L38,
  \dodoi{10.1088/2041-8205/793/2/L38}

\bibitem[{{Unno} {et~al.}(1989){Unno}, {Osaki}, {Ando}, {Saio}, \&
  {Shibahashi}}]{1989nos..book.....U}
{Unno}, W., {Osaki}, Y., {Ando}, H., {Saio}, H., \& {Shibahashi}, H. 1989,
  {Nonradial oscillations of stars} (Tokyo University Press, Tokyo)

\bibitem[{{Xu} \& {Lai}(2017)}]{2017PhRvD..96h3005X}
{Xu}, W., \& {Lai}, D. 2017, Phys. Rev. D, 96, 083005,
  \dodoi{10.1103/PhysRevD.96.083005}

\bibitem[{{Yoshida} \& {Lee}(2000{\natexlab{a}})}]{2000ApJS..129..353Y}
{Yoshida}, S., \& {Lee}, U. 2000{\natexlab{a}}, Ap. J. Suppl., 129, 353,
  \dodoi{10.1086/313410}

\bibitem[{{Yoshida} \& {Lee}(2000{\natexlab{b}})}]{2000ApJ...529..997Y}
---. 2000{\natexlab{b}}, Ap. J., 529, 997, \dodoi{10.1086/308312}

\bibitem[{{Yoshida} \& {Lee}(2002)}]{2002ApJ...567.1112Y}
---. 2002, Ap. J., 567, 1112, \dodoi{10.1086/338663}

\bibitem[{{Zaqarashvili} {et~al.}(2021){Zaqarashvili}, {Albekioni},
  {Ballester}, {Bekki}, {Biancofiore}, {Birch}, {Dikpati}, {Gizon},
  {Gurgenashvili}, {Heifetz}, {Lanza}, {McIntosh}, {Ofman}, {Oliver},
  {Proxauf}, {Umurhan}, \& {Yellin-Bergovoy}}]{2021SSRv..217...15Z}
{Zaqarashvili}, T.~V., {Albekioni}, M., {Ballester}, J.~L., {et~al.} 2021,
  Space. Sci. Rev., 217, 15, \dodoi{10.1007/s11214-021-00790-2}

\end{thebibliography}

\end{document}